# Constraints on LIGO/Virgo Compact Object Mergers from Late-time Radio Observations

Ashna Gulati,[1,2,3]⋆ Tara Murphy,[1,2] Dougal Dobie,[1,2,5] Adam Deller,[2,5] David L. Kaplan,[7] Emil Lenc,[3] Ilya Mandel,[6,2] Stefan Duchesne,[4] Vanessa Moss[3]

[1] *Sydney Institute for Astronomy, School of Physics, The University of Sydney, NSW 2006, Australia*
[2] *ARC Centre of Excellence for Gravitational Wave Discovery (OzGrav), Hawthorn, VIC 3122, Australia*
[3] *CSIRO Space and Astronomy, PO Box 76, Epping, NSW 1710, Australia*
[4] *CSIRO Space and Astronomy, PO Box 1130, Bentley WA 6102, Australia*
[5] *Centre for Astrophysics and Supercomputing, Swinburne University of Technology, Hawthorn, Victoria, Australia*
[6] *School of Physics and Astronomy, Monash University, Clayton, Victoria 3800, Australia*
[7] *Department of Physics, University of Wisconsin-Milwaukee, P.O. Box 413, Milwaukee, WI 53201, USA*

**ABSTRACT**
We present results from a search for radio afterglows of compact object mergers conducted with the Australian SKA Pathfinder. We used data from four epochs of the Rapid ASKAP Continuum Survey to search compact binary merger localization regions observed during the LIGO/Virgo O2, and O3 observing runs. Our investigation focused on eleven events (published in the GWTC-1, GWTC-2, and GWTC-3 catalogues of gravitational-wave events) with 90% posterior localisations smaller than 150 deg$^2$ and ≥99% probabilities of being of astrophysical origin, to identify potential radio afterglow-like transients up to ≲1500 days post-merger. We identified candidate afterglow-type variable sources in the 90% localisation for events– GW190503, GW200202 and GW200208, which were ruled out as unlikely to be related to the corresponding GW event on further analysis. Since we find no likely candidate counterparts, we constrain the inclination angle and the circum-merger density at isotropic equivalent energies ranging from $2 \times 10^{51} - 1 \times 10^{54}$ erg. These constraints are based on the assumption that the electron energy distribution in the associated jets follows a power-law index of $p = 2.2$, with 1% of the shock energy in the magnetic field ($\epsilon_B = 0.01$) and 10% in the electrons ($\epsilon_e = 0.1$). We discuss the detectability of late-time afterglows as a function of merger distance and inclination angles with millijansky surveys.

## 1 INTRODUCTION

The ground-based gravitational wave (GW) detectors Advanced Laser Interferometer Gravitational-Wave Observatory (LIGO) (LIGO Scientific Collaboration et al. 2015) and Advanced Virgo (Acernese et al. 2015) have played a big role in the development of multi-messenger astronomy. A notable breakthrough was the detection of a gravitational wave from a binary neutron star (NS) merger (Abbott et al. 2017a) on August 17, 2017, which was accompanied by radiation across the electromagnetic (EM) spectrum from gamma-rays to radio (Abbott et al. 2017c). Multi-wavelength observations were critical in studying the merger physics of this event in great detail. All four observed counterparts – the gamma-ray burst (Abbott et al. 2017d; Savchenko et al. 2017), the optical kilonova (Coulter et al. 2017; Tanvir et al. 2017), and the X-ray (Troja et al. 2017; Evans et al. 2017) and radio afterglows (Hallinan et al. 2017; Mooley et al. 2018b,a) – separately indicated the nature of the emission. The contradicting combination of weak gamma-ray emission suggesting slightly off-axis emission and a potentially delayed radio afterglow lightcurve with no sharp rise suggesting significant off-axis emission could only be explained by a structured jet wherein the Lorentz factor and outflow energetics vary with the angle from the core (Mooley et al. 2018a; Troja et al. 2017; Kasliwal et al. 2017; Hallinan et al. 2017). The identification of the merger's EM counterpart also allowed an estimate of the recession velocity which, in conjunction with the distance from the gravitational-wave event, allowed a measurement of the Hubble constant, independent of the cosmic distance ladder (Abbott et al. 2017b; Hotokezaka et al. 2019; Howlett & Davis 2020; Coughlin et al. 2020).

GW detectors can give us (i) the localisation of the sky location at which the merger happened by triangulating the arrival time of a GW signal at different detectors (Fairhurst 2009; Abbott et al. 2018) and (ii) the type of compact objects involved in the merger via template matching to the inspiral waveform (Baker et al. 2006). However, they do not provide information on the merger environment. Targeted electromagnetic follow-up of GW event localisations has enormous potential utility: it enables the identification of a host by looking for transients at optical wavelengths in possible host galaxies (Nuttall & Sutton 2010) immediately post-merger, and then following up at radio/X-ray wavelengths to reveal details about the merger environment and inclination angles.

---

⋆ E-mail: agul8829@uni.sydney.edu.au

[2] https://zenodo.org/records/6513631#.Y5gn4y0r2fU

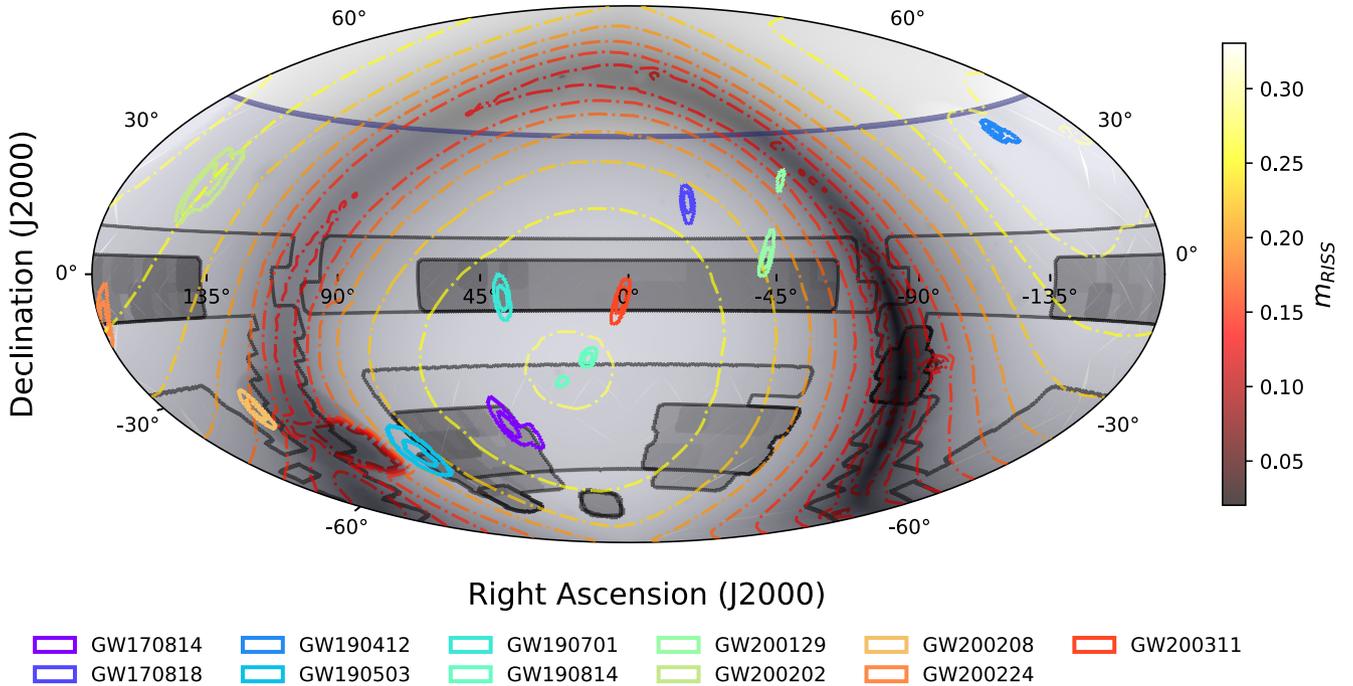

**Figure 1.** 50% and 90% posterior localisations of all GW events [2] included in this work for late-time afterglow search in equatorial coordinates in relation to the Rapid ASKAP Continuum Survey (RACS) and Variable and Slow Transients Survey (VAST) survey footprints used to search. The indigo solid line at a declination of 40° gives the northern limit for the RACS survey footprint covering the entire sky south of it. The black shaded regions give the VAST Pilot survey footprint and the black non-shaded regions give the VAST full survey footprint. The grey-shaded background shows the NE2001 electron density model map from the Cordes & Lazio (2002). The dot-dashed contours show the modulation index due to refractive interstellar scintillation at 888 MHz, calculated following Walker (1998).

Most models predict EM counterparts only when the compact object merger system has at least one NS (Berger 2014; Rezzolla et al. 2010; Giacomazzo et al. 2013), i.e., binary neutron star (BNS) or neutron star-black hole (NSBH) systems. This requirement arises from the need for tidally disrupted material from the NS to form an accretion disk around the newly formed BH post-merger, which supplies the required energy to fuel the jetted emission. However, only 6 out of 89 merger events detected in the first three LIGO/Virgo observing runs, O1, O2 and O3, with $\geq 99\%$ probability of being of astrophysical origin, fit this criterion and only two out of those are well-localised. For the 83 'clean' binary black hole mergers, no material is available to form an accretion disk and no associated EM signature is expected. Hence, binary black hole (BBH) mergers are not prime events for targeted follow-up. However, the reporting of potential EM counterparts for archival events, for instance: (i) the occurrence of a mildly significant gamma-ray trigger from *Fermi*-GBM lasting for 0.4 s after the BBH merger event GW150914 (Connaughton et al. 2016), and (ii) the spatial coincidence of optical flare ZTF 19abanrhr in an AGN within 18 days of GW190521 (Graham et al. 2020) has renewed interest in EM counterparts for BBH events that were expected to be electromagnetically dark.

These gamma-ray (Connaughton et al. 2016) and optical multiwavelength association claims (Graham et al. 2023) remain the subject of debate (Greiner et al. 2016; Connaughton et al. 2018; Lyutikov 2016; Xiong 2016; Ashton et al. 2021; Calderón Bustillo et al. 2021). These associations have prompted theoretical frameworks to explore the potential for EM emission via the launch of jetted emission powered by matter accretion (e.g., Janiuk et al. 2017; de Mink & King 2017; Graham et al. 2023; Tagawa et al. 2023), or alternate non-accretion mechanisms (e.g., Fraschetti 2018; Kelly et al. 2017) in a BBH system.

Compact object mergers are theorised to launch relativistic jets that interact with the circum-merger material, generating a shock front that relativistically accelerates electrons according to the standard fireball model (Mészáros & Rees 1997). The relativistic acceleration of electrons produces synchrotron radiation at radio frequencies, also known as the afterglow of the original jetted emission. We can test these theories proposing jetted emission similar to GRBs from BBH and NSBH mergers by searching for transients in our dataset that exhibit such afterglow-like emission as demonstrated in similar works by Dobie et al. (2022); Alexander et al. (2021); Bhakta et al. (2021). The typical isotropic equivalent kinetic energies for known jetted emission in short-GRBs are about $10^{49} - 10^{53}$ erg (Fong et al. 2015) and for long-GRBs are about $10^{49} - 10^{54}$ erg (Ghirlanda et al. 2009). Since no radio counterpart has been identified for a BBH or NSBH merger, we searched for off-axis afterglows from jets with all known jet energies, being model agnostic to GRB progenitors. Multi-epoch radio surveys are, perhaps, the only feasible way to search for off-axis afterglow-type EM counterparts from GW events, especially in the absence of targeted follow-up for BBH systems. Hence, we conduct a systematic search for radio counterparts to GW events using data from the Australian SKA Pathfinder (ASKAP).

In Section 2, we describe our search dataset and transient search. In Section 3, we discuss the results of the transient search. In section 4, we describe the constraints we can put on potential relativistic jets launched as a result of the compact object merger in archival GW events. In Section 5, we discuss the potential for finding EM counterparts and the implications of the constraints that millijansky-level surveys can put on the merger physics for gravitational-wave progenitors. In section 6, we conclude.

**Table 1.** Epochs of the Rapid ASKAP Continuum Survey used in the search and the observation specifications. Columns 1 through 8 give the epoch name, the dates it was observed over, the observing frequency, the declination limit up to which the observations were made, the area that the survey footprint covered, the integration time, the median root mean square (RMS) sensitivity and the reference.

| Survey | Dates Observed | Frequency (MHz) | $\delta$ Limit (°) | Area (deg$^2$) | $t_{int}$ (minutes) | Median RMS (mJy/ beam) | Reference |
|---|---|---|---|---|---|---|---|
| RACS-low | April 2019 - November 2020 | 887.5 | $\leq +41$ | ~34000 | 15 | 0.3 | McConnell et al. (2020) |
| RACS-mid | December 2020 - March 2021 | 1367.5 | $\leq +49$ | ~36000 | 15 | 0.2 | Duchesne et al. (2023) |
| RACS-high | December 2021 - Feb 2022 | 1655.5 | $\leq +48$ | ~36000 | 15 | 0.2 | Duchesne et al., in prep. |
| RACS-low2 | March 2022 - May 2022 | 887.5 | $\leq +49$ | ~36000 | 15 | 0.2 | — |

## 2 DATA AND SEARCH METHOD

### 2.1 Observational Data

ASKAP is a widefield radio telescope made up of an array of 36 prime focus antennas, located at Inyarrimanha Ilgari Bundara / the CSIRO Murchison Radio-astronomy Observatory in Western Australia (Johnston et al. 2007; Hotan et al. 2021), operating at frequencies between 700 and 1800 MHz. To investigate late-time afterglows, we used datasets from four epochs of the Rapid ASKAP Continuum survey (RACS; McConnell et al. 2020), covering the sky south of declination ~ 40°. The epochs included in the search are listed in Table 1. This included two epochs of RACS-low at 888 MHz (conducted in 2019 and 2022; McConnell et al. 2020), one epoch of RACS-mid at 1367 MHz (conducted in 2021, and covering a declination range of −80° to 41°; Duchesne et al. 2023), and one epoch of RACS-high at 1655 MHz (conducted in 2022, Duchesne et al., in prep.[3]). The RACS survey covers ~36,000 deg$^2$ with a typical root mean square (RMS) sensitivity of ~ 0.2 − 0.3 mJy for a ~ 15 minute integration at low, mid, and high observing bands. The point spread function for the RACS survey is declination dependent, ranging between 8.1″ − 47.5″ for the RACS-mid epoch; and from 13.2″ - 45.1″ for the RACS-low epochs as described in Fig. 4 of McConnell et al. (2020). We also incorporated data from the ASKAP Variable and Slow Transients Pilot Survey (VAST; Murphy et al. 2021) and 10 epochs of ASKAP observations for the GW190814 field from Dobie et al. (2022). The VAST pilot survey covers approximately ~ 5,000 deg$^2$ with a typical RMS sensitivity of ~ 0.3 mJy for a ~ 12 − 15 minute integration at low bands, while the GW190814 observations have an RMS sensitivity of 35 $\mu$Jy (10-hour integration time). These additional datasets were essential for augmenting our observational coverage, particularly when the footprint coincided with the localization region of the GW merger events.

In our search, we have included significant gravitational wave events included in the GWTC 1 (Abbott et al. 2019), GWTC 2.1 (Abbott et al. 2021) and GWTC 3 (Abbott et al. 2023) catalogues. These catalogues contain events observed during the LIGO/Virgo ground-based gravitational wave detector network runs — O1, O2 and O3. We selected the events that had a 90% posterior localisation within the RACS survey region and ≥99% probability of being of astrophysical origin. From those we selected, we chose events that had 90% localisations of $\leq 150$ deg$^2$ (see Fig. 1) and no confirmed EM counterparts.

### 2.2 Search Method

We searched for afterglow-like transients associated with 11 confident well-localised gravitational-wave events (see Table 2). We used vast-tools (Stewart et al. 2023) to get the number of sources from the two RACS-low epochs within each event localisation, given in Table 2. We included sources that:

(i) are compact, as defined in Hale et al. (2021) by the integrated-to-peak flux-density ratio: $S_I/S_P < 1.025 + 0.69 \times SNR^{-0.62}$, where peak flux corresponds to the pixel with the maximum flux density value and integrated flux corresponds to the sum total of flux densities in all pixels covering the source;

(ii) have a SNR $\geq 10$ for sources with one detection and SNR $\geq 7.5$ for sources with more than one detection (see Metzger et al. 2015).

We then identified variable sources within each of the datasets. We used the two RACS-low epochs at 888 MHz, in conjunction with scaled RACS-mid (1367 MHz) and RACS-high (1650 MHz) data, and computed modulation indices for sources with detections occurring after the date of the merger. We scaled the RACS-mid and RACS-high flux density measurements to 888 MHz to match RACS-low, assuming a spectral index of $\alpha = -0.6$ ($\frac{p-1}{2}$) as expected for synchrotron emission assuming an electron energy distribution with a power law index of 2.2 (Condon & Ransom 2016). We use 888 MHz as the reference frequency as the flux density of RACS-low is well characterised in McConnell et al. (2020) compared to the mid/high bands.

Refractive interstellar scintillation (RISS) is the scattering of plane wavefronts travelling from a source to an observer on Earth after passing through large-scale inhomogeneities in the interstellar medium (ISM) (see review by Hancock et al. 2019). RISS causes compact radio sources to exhibit significant variability and manifest as false positives in radio synchrotron transient searches. Following the procedure of Dobie et al. (2022) we estimate the expected RISS-induced variability for a point source at our observing frequency using the NE2001 electron density model by Cordes & Lazio (2002) and the equations of Walker (1998). We identified a source as variable if the source's modulation index exceeded the expected averaged modulation index within the localisation region due to RISS (given in column (2) Table 3).

For sources with two detections, we applied the modulation index definition from Mooley et al. (2016), while for those with more than two detections, we followed the definition in Mooley et al. (2013), given by :

---

[3] https://research.csiro.au/racs/racs-high1-dr1-raw-data/

[4] The event GW200208 refers to the GW200208_130117 event.

Table 2. Description of the gravitational-wave events included in this search. Columns 1 through 9 give the name of the event, the LIGO/Virgo run it was detected in, its distance (here and elsewhere, the median and 90% credible interval quantiles are quoted from the catalogues), the masses of the compact objects involved in the merger, the inclination angle, radiated energy (defined as the difference between the source total and source remnant mass), the 90% posterior localisation and the median number of days post-merger it was observed with ASKAP. The 90% credible interval for the inclination angles and radiated energy has been deduced from the gravitational wave signal parameter estimation using the mixed set of samples created by fitting both waveforms IMRPhenomXPHM and SEOBNRv4PHM.

| GW Name | Detector Run | Distance (Mpc) | Mass 1 ($M_\odot$) | Mass 2 ($M_\odot$) | $\theta_{obs}$ (°) | $E_{GW}$ ($M_\odot$) | 90% localisation (deg$^2$) | $\Delta T$ (days) |
|---|---|---|---|---|---|---|---|---|
| (1) | (2) | (3) | (4) | (5) | (6) | (7) | (8) | (9) |
| GW170814 | O2 | $600^{+150}_{-220}$ | $30.6^{+5.6}_{-3.0}$ | $25.2^{+2.8}_{-4.0}$ | $40^{+108}_{-28}$ | $2.8^{+0.4}_{-0.4}$ | 80 | [630, 1254, 1623, 1712] |
| GW170818 | O2 | $1060^{+420}_{-380}$ | $35.4^{+7.5}_{-4.7}$ | $26.7^{+4.3}_{-5.2}$ | $37^{+31}_{-27}$ | $2.9^{+0.6}_{-0.5}$ | 27 | [614, 1234, 1600, 1700] |
| GW190412 | O3 | $720^{+240}_{-220}$ | $27.7^{+6.0}_{-6.0}$ | $9.0^{+2.0}_{-1.4}$ | $40^{+17}_{-14}$ | $1.2^{+0.2}_{-0.2}$ | 25 | [9, 622, 996, 1076] |
| GW190503 | O3 | $1520^{+630}_{-600}$ | $41.3^{+10.3}_{-7.7}$ | $28.3^{+7.5}_{-9.2}$ | $36^{+38}_{-26}$ | $3.00^{+0.95}_{-1.07}$ | 103 | [1, 634, 992, 1081] |
| GW190701 | O3 | $2090^{+770}_{-740}$ | $54.1^{+12.6}_{-8.0}$ | $40.5^{+8.7}_{-12.1}$ | $32^{+30}_{-23}$ | $4.1^{+1.1}_{-1.2}$ | 44 | [558, 926, 1017] |
| GW190814 | O3 | $230^{+40}_{-50}$ | $23.3^{+1.4}_{-1.4}$ | $2.6^{+0.1}_{-0.1}$ | $45^{+17}_{-11}$ | $0.24^{+0.01}_{-0.01}$ | 22 | [501, 871, 975] |
| GW200129 | O3 | $900^{+290}_{-380}$ | $34.5^{+9.9}_{-3.2}$ | $28.9^{+3.4}_{-9.3}$ | $37^{+35}_{-28}$ | $3.2^{+0.4}_{-0.9}$ | 31 | [358, 712, 803] |
| GW200202 | O3 | $410^{+150}_{-160}$ | $10.1^{+3.5}_{-1.4}$ | $7.3^{+1.1}_{-1.7}$ | $32^{+33}_{-24}$ | $0.8^{+0.1}_{-0.1}$ | 150 | [337, 696, 782] |
| GW200208 [4] | O3 | $2230^{+1000}_{-850}$ | $37.8^{+9.2}_{-6.2}$ | $27.4^{+6.1}_{-7.4}$ | $36^{+36}_{-26}$ | $2.8^{+0.8}_{-0.8}$ | 30 | [347, 711, 793] |
| GW200224 | O3 | $1710^{+490}_{-640}$ | $40.0^{+6.9}_{-4.5}$ | $32.5^{+5.0}_{-7.2}$ | $36^{+33}_{-26}$ | $3.6^{+0.7}_{-0.7}$ | 42 | [315, 680, 764] |
| GW200311 | O3 | $1170^{+280}_{-400}$ | $34.2^{+6.4}_{-3.8}$ | $27.7^{+4.1}_{-5.9}$ | $31^{+29}_{-22}$ | $2.9^{+0.5}_{-0.6}$ | 35 | [292, 681, 769] |

$$m = \begin{cases} \frac{\Delta S}{\overline{S}}, & \text{if } N_{det} = 2 \\ \frac{1}{\overline{S}} \sqrt{\frac{1}{N-1} \sum_i^N (S_i - \overline{S})^2}, & N_{det} > 2 \end{cases} \quad (1)$$

where $S$ denotes flux density, $\Delta S = |S_2 - S_1|$, $\overline{S}$ is the mean flux density and $S_i$ is the flux density for each detection of the source. We also assessed whether the t-statistic for the source being variable exceeds the 95% threshold, as described in Mooley et al. (2016), given by:

$$V_s = \frac{\Delta S}{\sqrt{\sigma_1^2 + \sigma_2^2}} \geq 4.3, \quad (2)$$

where $\sigma_1$ and $\sigma_2$ are the standard errors on the flux density. For sources with only one detection across the four epochs, we selected those with a fractional flux density change from the mean non-detection $5\sigma$ limit, as expected from the modulation index (see Mooley et al. 2016).

The sample of variable sources not consistent with scintillation, i.e. those sources with a modulation index greater than $m_{RISS}$, are still contaminated by different kinds of radio objects that can vary either intrinsically or due to scintillation such as pulsars, flaring stars, scintillating AGN and other synchrotron transients, prevalent at GHz frequency surveys (see discussion in Murphy et al. 2021; McConnell et al. 2020). We filtered these sources out using several cuts as follows:

(i) We excluded sources with circular polarisation within a 1″ radius, typically associated with coherent emission processes in objects like flare stars (e.g., Pritchard et al. 2021) and pulsars (e.g., Lenc et al. 2018). In the context of synchrotron radiation, circular polarization is generally very weak (Macquart 2002; Corsi et al. 2018), and so we can exclude sources with significant circular polarisation. Since less than 0.03% of the RACS radio sources in this sample have genuine circular polarisation, there are no sources eliminated in this cut.

(ii) We excluded sources that were already bright pre-merger, indicating that they are unrelated to the compact binary merger. It should be noted that this cut will discard any sources that are in a galaxy with detectable radio emission due to star formation, or AGN. We excluded these by cross-matching our data with archival radio surveys – the National Radio Astronomy Observatory VLA Sky Survey (NVSS; Condon et al. 1998), the Sydney University Molonglo Sky Survey (SUMSS; Mauch et al. 2003), the second epoch Molonglo Galactic Plane Survey (MGPS-II; Murphy et al. 2007), the VLA Faint Images of the Radio Sky (FIRST; Becker et al. 1995), the Galactic and Extra-Galactic All-Sky MWA (GLEAM; Hurley-Walker et al. 2017) and the MeerKAT Galaxy Cluster Legacy Survey (MGCLS; Knowles et al. 2022) using a crossmatch within a 3″ radius and removed the remaining using manual checks. The number of sources remaining after this cut is recorded for each event under column (5) in Table 3.

(iii) We removed sources with a modulation index less than $m_{RISS}$ in subsequent VAST observations (accessed via CASDA in February 2024). The number of sources remaining after this cut is recorded for each event under column (6) in Table 3.

(iv) We excluded Widefield-Infrared Survey Explorer (WISE) sources within 5″, wherein the WISE colours were indicative of it either being a star or AGN. This determination was made using the WISE [3.4 µm] − [4.6 µm] and [4.6 µm] − [12 µm] infrared color-color classification system introduced by Wright et al. (2010). When colours were found consistent with either classification, we eliminated only sources that correspondingly had a > 99% class probability in the GAIA DR3 catalog (Gaia Collaboration et al. 2023), rendering a false classification fraction of < 1%. We removed known AGN because given our limited sampling, we would not be able to target EM emission scenarios involving mergers in AGN disks, as studied in Graham et al. (2023). We also removed previously known sources, determined through crossmatching with the Gaia catalogue (Gaia Collaboration et al. 2023) within 2″, pulsar scraper (Kaplan 2022) within 2″, and transient name server[5] within 10″. The number of sources remaining after this cut is recorded for each event under column (7) in Table 3.

Table 3. Summary of late-time candidate search. Column 1 gives the event name; Column 2 gives the modulation index expected from refractive interstellar scintillation in this region from the NE2001 electron density model, $m_{RISS}$; Column 3 gives the number of RACS sources in the event's 90% localisation; Column 4 gives the number of compact sources above the modulation index cut; Columns 5 through 10 give the number of sources remaining after removing sources from the variable sample: with archival radio detections, with modulation index $\leq m_{RISS}$ in additional ASKAP monitoring, that are known, that have artefact variability, that have lightcurves spectrally or temporally inconsistent with an afterglow and those that could not be ruled out based on prior criteria. Sources surviving all cuts, and thus appearing in the last column, are candidate detections.

| | | Number of Sources | | | | | | | |
|---|---|---|---|---|---|---|---|---|---|
| GW Name | $m_{RISS}$ | RACS | Variable | Archival Radio | ASKAP Non-variable | Known | Artefact Variability | Afterglow Inconsistent | Candidates |
| (1) | (2) | (3) | (4) | (5) | (6) | (7) | (8) | (9) | (10) |
| GW170814 | 0.29 | 14297 | 324 | 188 | 90 | 72 | 52 | 0 | 0 |
| GW170818 | 0.26 | 3779 | 112 | 46 | 46 | 44 | 44 | 0 | 0 |
| GW190412 | 0.32 | 4226 | 75 | 2 | 2 | 2 | 1 | 0 | 0 |
| GW190503 | 0.25 | 20156 | 469 | 227 | 184 | 162 | 155 | 2 | 2 |
| GW190701 | 0.29 | 6370 | 20 | 3 | 3 | 2 | 0 | 0 | 0 |
| GW190814 | 0.30 | 3679 | 21 | 16 | 3 | 3 | 2 | 0 | 0 |
| GW200129 | 0.24 | 4554 | 22 | 4 | 3 | 3 | 2 | 0 | 0 |
| GW200202 | 0.28 | 23764 | 61 | 3 | 3 | 3 | 2 | 1 | 1 |
| GW200208 | 0.19 | 5924 | 37 | 12 | 12 | 10 | 3 | 1 | 1 |
| GW200224 | 0.28 | 7034 | 33 | 11 | 4 | 4 | 0 | 0 | 0 |
| GW200311 | 0.30 | 5574 | 24 | 6 | 4 | 2 | 0 | 0 | 0 |

(v) We removed artefacts associated with poor image quality and source finder errors. The number of sources remaining after this cut is recorded for each event under column (8) in Table 3.

(vi) We used spectral and temporal information from additional radio data (VLASS, accessed via CIRADA[6] and VAST, accessed via CASDA in February 2024) and made the following cuts to exclude sources that are not consistent with the spectral and temporal behaviour expected of afterglows:

• Sources that did not exhibit the expected rise and fall pattern of afterglows i.e. were persistent, rising after falling or stochastically varying in follow-up observations. This included a manual vetting process.

• Sources with a spectral index (fitted across the four RACS epochs) that was flat or inverted (fitted $3\sigma$ lower bound spectral index $\geq 0$ with $R^2 \geq 0.8$). This cut only eliminates well-fit, non-variable, inverted-spectrum sources, making it an effective method for identifying and removing AGN, which constitute the majority of contaminating sources in this search. It removes few true afterglows, as afterglow emission at late times is typically optically thin ($\alpha < -0.5$; Dobie et al. 2018) and time-varying, meaning it is rarely well described by a constant flat or inverted spectrum over time.

We investigate the number of true afterglows that may be removed by these cuts (including the effect of synchrotron self-absorption, which is not modelled by afterglowpy) in Section 4.3. The number of sources remaining after this cut is recorded for each event under column (9) in Table 3.

For each remaining afterglow-like transient, we employ Markov chain Monte Carlo (MCMC) fitting, using the Python package emcee (Foreman-Mackey et al. 2013). We used afterglowpy[7] (Ryan et al. 2020) to fit a tophat jet in which all of the jet energy is contained within a narrow cone with an opening angle and a structured Gaussian jet with a relativistic core surrounded by Gaussian wings of slower-moving material (see Ryan et al. 2020, for a detailed discussion). We fitted off-axis afterglows because, aside from the speculative EM counterparts reported in temporal and spatial coincidence for two archival GW events, GW150914 and GW190521, no such coincidence has been reported for any of the other events. This suggests either that there was no jet, or the jet was not directly observable in our line of sight. It is also possible the gamma-ray detectors were not on or not pointing in the right direction, however, we assume optimal detector functionality for this search.

The flat priors on the following parameters used for fitting are defined as follows: isotropic equivalent kinetic energy, $E_{K,iso} \in \{1 \times 10^{49}, 1 \times 10^{54}\}$ erg; jet half-opening angle, $\theta_{j/2} \in \{0 - 20°\}$; circum-merger density, $n \in \{0.001 - 10\}$ cm$^{-3}$, and the wing span or truncation angle outside of which the energy is initially zero for a Gaussian jet, $\theta_{wing} \in \{20 - 89°\}$. Here we follow afterglowpy in using the jet half-opening angle. We used Gaussian priors for the viewing angle[8], $\theta_{obs}$ and distance luminosity, $d_L$. The mean and standard deviation for the Gaussian were deduced from each event's GW signal parameter estimation posteriors taken from the mixed set of samples created by fitting both waveforms `IMRPhenomXPHM` and `SEOBNRv4PHM`. The bounds on the viewing angle were set by the GWTC catalogue's posterior probability estimations with the condition that the viewing angle has to be greater than the half-opening angle to fit sources with off-axis afterglow lightcurves only. The limits for the luminosity distance, $d_L$ were set based on the GWTC catalogue's posterior probability estimations given in Table 2. We assume the microphysical parameters $p = 2.2$, $\epsilon_e = 0.1$, and $\epsilon_B = 0.01$, consistent with previous studies (e.g., Troja et al. 2017; Alexander et al. 2021). Including these as variables would introduce too many parameters to constrain with our limited observations. We assume a flat lambda-CDM cosmology with $H_0 = 67.8$ km s$^{-1}$ Mpc$^{-1}$ (Planck Collaboration et al. 2016). We chose 500 walkers for tophat model fittings and 600 walkers for Gaussian model fittings, with 3000 steps each. We obtained the best-fit values and the 68% confidence interval from the posterior after removing burn-in of up to the first 500 steps. In cases of nominally good fits to the trend of the data, we also cross-checked against the NASA Extragalactic database[9] for counterpart galaxies within the expected physical offsets for short-GRBs of $\leq 30$ kpc (Fong et al. 2010). We consider a model to be a nominally good fit to the data if the $\chi^2$ value ($\chi^2 = \sum \frac{(S_{\rm expected} - S_{\rm observed})^2}{S_{\rm error}^2}$) is low and accounts for any non-detections. The goodness of fit is determined using the $\chi^2$ diagnostic for cases where the number of detections, $n_{\rm det} \leq 6$. For $n_{\rm det} > 6$, we use the reduced $\chi^2$ diagnostic ($\chi^2_\nu = \frac{\chi^2}{n_{\rm dof}}$, where $n_{\rm dof} = n_{\rm det} - n_{\rm model\_params}$). This approach is necessary because our models involve five (tophat jet models) or six (Gaussian jet models) parameters, and using the reduced $\chi^2$ when the number of detections is fewer than the number of parameters would lead to inconclusive values. We also check the fit's explanation for non-detections by calculating the $\chi^2$ for all datapoints where a non-detection flux was taken as $5\sigma$ of the RMS value.

---

[5] https://www.wis-tns.org
[6] https://cirada.ca/
[7] We note that afterglowpy (version 0.8.0; https://github.com/geoffryan/afterglowpy) does not include synchrotron self-absorption, which may affect model fitting in the radio band. It does allow modeling for off-axis angles, but for very large angles, the jet emission profile and relativistic beaming effects may not be fully represented.
[8] Viewing angle is interchangeably used with inclination angle throughout.

**Table 4.** Summary of tophat and Gaussian jet model fits to the candidates. Column 1 gives the event name; Column 2 gives the candidate number; Columns 3 and 4 give the right ascension and declination of the candidate; Column 5 gives the number of detections we use to fit the model; Columns 6 and 7 give the $\chi^2$ for the models fit.

| GW Name | Candidate | Candidate RA | Candidate Dec | $n_{det}$ | TopHat Fit $\chi^2$ | Gaussian Fit $\chi^2$ | $L_{\rm radio, 888MHz}$ (erg s$^{-1}$) |
|---|---|---|---|---|---|---|---|
| GW190503 | 1 | 6:12:48.5321 | -50:45:34.0272 | 4 | 57 | 58 | $\sim 10^{39}$ |
|  | 2 | 6:33:07.6286 | -47:00:14.7132 | 5 | 65 | 37 | $\sim 10^{39}$ |
| GW200202 | 1 | 9:16:43.0920 | 29:48:11.3818 | 3 | 22 | 22 | $\sim 10^{38}$ |
| GW200208 | 1 | 9:13:21.3482 | -36:04:26.5116 | 7 | 54 | 64 | $\sim 10^{40}$ |

## 3 SEARCH RESULTS

We found four candidates that could not excluded by any of our filtering criteria. Two for the event GW190503, and one each for events GW200202 and GW200208. These candidates are given in Table 4 and the model fits are given in Figure 2 (the rest of the figures are given in Figures 4 to 7 of online-only supplementary material). In this section, we discuss these candidates. We found no afterglow candidates for 8 out of 11 of our events. We discuss the observational constraints that we can place on these non-detections with our survey sensitivity in Section 4.

[9] https://ned.ipac.caltech.edu

### 3.1 BBH GW190503

GW190503 was reported as a binary black hole merger with component masses of $41.3^{+10.3}_{-7.7}$ M$_\odot$ and $28.3^{+7.5}_{-9.2}$ M$_\odot$ at a luminosity distance of $1520^{+630}_{-600}$ Mpc (Abbott et al. 2021). We searched 90% of the final GWTC-2.1 catalogue map (an area of 102.74 deg$^2$). We had four post-merger observations at 1, 634, 992, and 1081 days at 888, 1367, 1655 and 888 MHz respectively. Of the 20 156 sources in this region, 492 were compact and above our 0.25 modulation index cut. After filtering out sources with archival radio detections, known sources, and those showing artefact variability, two potential transient sources remained.

These potential candidates are at a location in the 28.4 and 87.9-percentile of the GW190503 localisation respectively. Both these potential candidates brightened post-merger (1 and 1.4 mJy 5-sigma upper limits on May 4, 2019, 1 day post merger, at 888 MHz), compliant with the criteria outlined in Section 2. They poorly fit the off-axis afterglow models, with $\chi^2$ values greater than 50 for tophat jets and greater than 35 for Gaussian jets (see Fig. 2 of online-only supplementary material). This is because the spectra are inconsistent with the expected SED for a synchrotron afterglow under the assumption of a reasonably smooth light curve evolution for multiple epochs. Additionally, the posterior sample shapes push against the prior boundaries, maximizing isotropic energy and core angle while minimizing luminosity distance. Hence, these are unlikely to be related to the gravitational-wave event. Assuming the source is unrelated to this event, our observational constraints are given in Table 5.

### 3.2 BBH GW200202

GW200202 was reported as a binary black hole merger with component masses $10.1^{+3.5}_{-1.4}$ M$_\odot$ and $7.3^{+1.1}_{-1.7}$ M$_\odot$ at a luminosity distance of $410^{+150}_{-160}$ Mpc (Abbott et al. 2023). Our search covered 90% of the final GWTC-3 catalogue map (an area of 150.31 deg$^2$). We had three post-merger observations, at 337, 696, and 782 days, at 1367 MHz, 1655 MHz, and 888 MHz, respectively. A pre-merger observation at 888 MHz enabled us to eliminate potential counterparts that were already bright pre-merger. Of the 23 764 sources in this region, 69 were compact and above our 0.28 modulation index cut. After filtering out sources with archival radio detections, known sources, and those showing artefact variability, only one potential candidate remained.

This potential candidate is at a location in the 89.6-percentile of the GW200202 localisation, brightening post-merger (0.7 mJy 5 sigma upper limit on May 20, 2019, VLASS 3 GHz), compliant with criteria outlined in Section 2. The fit for the off-axis afterglow models for a tophat and a Gaussian jet has a $\chi^2$ of 22 (see Fig. 2 of online-only supplementary material). However, there is a WISE infrared source 1.6″ away, which is a known bright galaxy at a redshift of 0.22. The redshift of this galaxy lies outside the 95% credible redshift range allowed by the GW posteriors. Hence, this transient is unlikely to be related to the gravitational-wave event. Assuming the source is unrelated to this event, our observational constraints are given in Table 5.

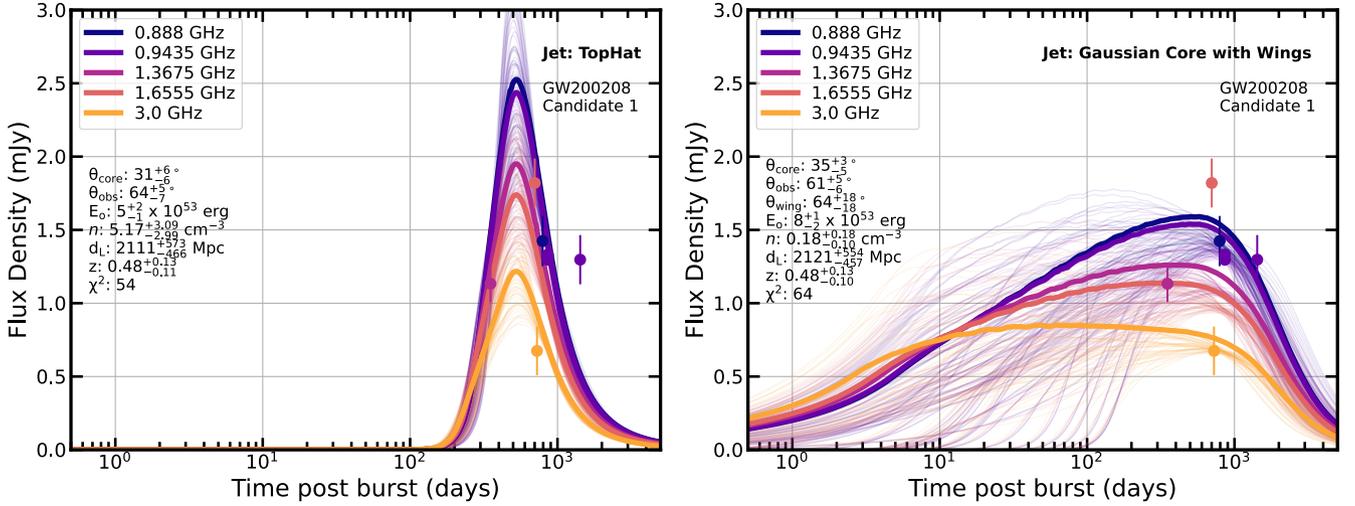

**Figure 2.** `Afterglowpy` Tophat ($\chi^2_\nu = 27$) and Gaussian jet fit ($\chi^2_\nu = 64$) to the radio data at a location in the 79.5-percentile of the GW200208 localisation. The observed radio flux densities are plotted in circles and 50 random MCMC samples are plotted in solid lines to demonstrate the uncertainty in the fits. The effect of RISS (which could introduce variability at the 19% level for this localisation) is not included in these uncertainties. The median and corresponding 68% credible interval fit parameters are given in the bottom left corner. See Fig. 3 of online-only supplementary material for posteriors from the MCMC analysis.

### 3.3 BBH GW200208

GW200208 was reported as a binary black hole merger with component masses of $37.8^{+9.2}_{-6.2}$ M$_\odot$ and $27.4^{+6.1}_{-7.4}$ M$_\odot$ at a luminosity distance of $2230^{+1000}_{-850}$ Mpc (Abbott et al. 2023). We searched 90% of the final GWTC-3 catalogue map (an area of 30.43 deg$^2$). We had three post-merger observations at 347, 711, and 793 days at 1367 MHz, 1655 MHz, and 888 MHz respectively. There was also a pre-merger observation at 888 MHz that allowed us to rule out potential counterparts that were already bright pre-merger. Of the 5 924 RACS sources in this region, 38 were compact and above our 0.19 modulation index cut. After filtering out sources with archival radio detections, known sources, and those showing artefact variability, only one potential candidate remained.

This potential candidate is at a location in the 79.5-percentile of the GW200208 localisation, brightening post-merger (0.9 mJy 5 sigma upper limit on July 7, 2019, VLASS 3 GHz), compliant with criteria outlined in Section 2. The off-axis afterglow models for tophat jets ($\chi^2_\nu = 27$) fit better than Gaussian jets ($\chi^2_\nu = 64$) (as shown in Fig. 2). Some datapoints are not so well explained by the model and there may be a systematic error in the survey data for that epoch. Further analysis shows the candidate has an isotropic luminosity of $\sim 10^{40}$ erg s$^{-1}$ at a lightcurve peak of 1.5 mJy at 888 MHz, typical of known radio afterglows. There is a WISE infrared source 3.4" away, with colours suggestive of a galaxy (see Fig. 2 of online-only supplementary material), but its distance remains unknown. If the WISE source is within the GW distance range, the candidate's offset would be 22–54 kpc which overlaps with the range for typical afterglow offsets. Hence, we cannot conclusively rule out this candidate based on poorly fit models; follow-up and analysis are required to determine whether this source could be linked to a structured jet from the event.

However, given the lower modulation index threshold of 0.25 for this region implies a high false positive rate for the identification of a source as a candidate. Hence, it is unlikely to be related to the gravitational-wave event. Assuming the source is unrelated to this event, our observational constraints are given in Table 5.

## 4 CONSTRAINTS ON RELATIVISTIC JETS LAUNCHED BY COMPACT OBJECT MERGERS

We found potential afterglow candidates in the 90% posterior probability region for three of the 11 events, but none are likely associated with the GW events. No candidates were found for the remaining eight events. In this section, we will discuss the observational constraints that this absence of afterglow-like transients in our search allows on the environments and energetics of the relativistic jets that may have been launched by these compact object mergers.

As we found no counterparts, we can rule out combinations of circum-merger density and viewing angle that would give afterglows brighter than the 5$\sigma$ detection limit of $\sim 1.5$ mJy from our ASKAP observations at the relevant post-merger times. These sensitivity constraints are given in Table 5. Note that multiple RACS images, observed over a few months, need to be considered for each GW event, to cover the 90% sky localisation. Hence we use the mode number of days post-merger for each group of observations to interpret the observational constraints for each event. The range of observation times is typically within 10% of the time post-merger. Hence, this does not significantly affect the results as the afterglows evolve on roughly logarithmic timescales.

We used `afterglowpy` (Ryan et al. 2020) to generate a grid of light curves at 888 MHz for two relativistic jets: a tophat jet with a jet opening angle, $\theta_j$, of 5° and 15°; and a Gaussian jet with a jet core $\theta_j$ of 5° and 15° and wings extending to four times the jet half opening angle i.e. 10° and 30° respectively. We assume that the compact object merger—whether BBH or BHNS—is associated with jetted emission, being agnostic to the jet-progenitor. Therefore, instead of using sGRB-specific opening angles, we adopt more general values: a typical opening angle of 5° (Salafia et al. 2023) and an optimistic wider angle of 15°. We generated these grids at isotropic-equivalent kinetic energies, $E_{K,iso}$, of $2 \times 10^{51}$, $5 \times 10^{51}$, $5 \times 10^{52}$ and $1 \times 10^{54}$ erg. We chose these energies to cover the wider range of energies for jetted emission in GRBs (Ghirlanda et al. 2009) and to ensure comparability with prior studies (Alexander et al. 2021; Dobie et al. 2022). We computed the light curves for off-axis viewing angles ranging from $2.6° - 90°$ and $7.6° - 90°$ for jets with $2.5°$ and $7.5°$ half-opening angles respectively and fixed circum-merger densities from $10^{-3}$ to $10$ cm$^{-3}$.

When the external forward shock interacts with the surrounding medium, the electrons are accelerated to relativistic speeds. At this point, the distribution of electrons with a Lorentz factor $\gamma_e$ is assumed to scale with $\gamma_e^{-p}$, where $p$ is the power-law index of the electron energy distribution (Santana et al. 2014). The electrons at the forward shock are in the fast-cooling regime (Piran 2004), hence, we fixed $p = 2.2$ (Tak et al. 2019). The microphysical parameters $\epsilon_B$, which describes the fraction of energy that is in the magnetic field downstream of the shock front, and $\epsilon_e$, the fraction of energy in the electrons in the shocked fluid, are taken to be constant throughout the afterglow emission. We used $\epsilon_e = 0.1$ (Santana et al. 2014; Sironi & Spitkovsky 2011). While $\epsilon_B$ could range over five orders of magnitude (Santana et al. 2014), we use the typical $\epsilon_B$ of 0.01 from previous studies (e.g., Troja et al. 2017; Alexander et al. 2021). Our constraints are a function of these microphysical parameters. We find that for our models, the afterglow flux density proportionally increases by an order of magnitude with $\epsilon_B$. Hence, using a higher value of 0.1 would lead to more stringent constraints as opposed to a lower value of 0.0001. We assumed an electron acceleration fraction of 100%. We set the uncertainties on the distance according to the GW event distance luminosity posterior estimates from the GWTC catalogues.

We can only rule out combinations of circum-merger density and viewing angles that produce detectable afterglows if the afterglows have modulation indices greater than that expected from scintillation. We therefore calculate the modulation index for detectable off-axis afterglows for each combination of circum-merger density and viewing angles and rule out only those combinations wherein $m \geq m_{RISS}$. These scintillation constraints are given in Table 5. Since we are not sensitive to BBH mergers that occur in AGN disks (Graham et al. 2023; Alves et al. 2024) we do not discuss constraints on such mergers with their very different physical environments.

Our sensitivity and scintillation constraints are summarised in Table 5. In the table, the lower limit on circum-merger density is given relative to the viewing angle limits from the GW data. For example, in the case of GW190412, if a jet with a $15°$ Gaussian core was launched at an isotropic equivalent energy of $5 \times 10^{52}$ erg, the minimum density (n) that can be ruled out at the lower bound of the viewing angles allowed by the GW analysis is 0.05 cm$^{-3}$. While lower circum-merger densities are possible (see Figures 4-7 in online-only supplementary material), the minimum viewing angle allowed by the GW detector posteriors is $27°$, thus it is noted as $\geq 0.05$, 27 in Table 5. For $E_{K,iso}$ of $1 \times 10^{54}$ erg, the minimum n is 0.001 cm$^{-3}$ at a viewing angle of $36°$, and this minimum extends to viewing angles lower than where it first occurs, noted as $\geq 0.001$, $\leq 36$ in Table 5. For GW170818, with a $15°$ tophat jet and $E_{K,iso}$ of $1 \times 10^{54}$ erg, the minimum n is 0.001 cm$^{-3}$ at $22°$, but this minimum does not extend to viewing angles below $22°$ (see Figures 4-7 in online-only supplementary material), so it is noted as $\geq 0.001$, 22 in Table 5. For GW200202, with a $15°$ tophat jet and $E_{K,iso}$ of $5 \times 10^{52}$ erg, the minimum n extends toward larger viewing angles, noted as $\geq 4$, $\geq 70$ (see Figures 4-7 in online-only supplementary material). However, since this is outside the viewing angles allowed by the GW detector, it is not included in the table. Figure 3 and online-only supplementary material show the constraints for all events included in our search. In the constraint plots, exclusion regions are marked based on the lower and upper bounds of the distance luminosity posterior from the GW detector data for different jet energies. These constraints allow us to exclude regions of the parameter space with higher circum-merger densities, which would produce brighter afterglows detectable by ASKAP.

### 4.1 Sensitivity and Scintillation Constraints

The observations and analyses of 10 binary black hole (BBH) mergers and one potential neutron star-black hole (NSBH) merger GW190814, led to the following constraints.

#### 4.1.1 Events with constraints on jets for $\theta_j$ of $5°$ and $15°$ for $E_{K,iso} \lesssim 5 \times 10^{52}$ erg

Our observations imposed stringent constraints on the circum-merger environment and potential jet properties for events: GW190412, GW190503, GW190814 and GW200202. The sensitivity of the radio observations allowed us to exclude regions for jet models with isotropic kinetic energies $\lesssim 5 \times 10^{52}$ erg at circum-merger densities ranging $0.001 - 7$ cm$^{-3}$. The average $m_{RISS}$ in these event regions ranging from 0.25–0.32 allowed constraints similar to the sensitivity constraints. The constraints were dependent on the assumed jet morphology — core opening angle, jet structure, and energy. Tophat and Gaussian jets with larger opening angles (for a given isotropic equivalent energy) and higher isotropic equivalent energies allowed a greater portion of the parameter space to be excluded as such events would have greater total energy and produce brighter afterglows. On the other hand, smaller opening angles or lower energies led to less stringent constraints.

Despite the large distance for events GW190412 ($d_L$ lower bound 500 Mpc) and GW190503 ($d_L$ lower bound 920 Mpc), the constraints are stringent because the observations are at early times ($1 - 10$ days) post-merger (see Fig. 4). We note that afterglows for tophat jets are less constrained for GW190412 compared to Gaussian jets as the viewing angles allowed by the GW data are significantly off-axis ranging $26° - 57°$. With our choice of normalisation, tophat jets have less total energy than Gaussian jets for the same on-axis isotropic equivalent energy and core opening angle, so they rise and fall faster (see Fig. 5), and would have lower flux densities compared to a Gaussian jet for the same viewing angle.

For events GW190814 and GW200202, these stringent constraints are a result of lower distance (lower bound on $d_L \leq 250$ Mpc). Our late-time observations exclude greater regions for larger viewing angles as compared to smaller ones. For GW190814, our constraints are complementary to those determined from earlier observations conducted by Dobie et al. (2019); Alexander et al. (2021); Dobie et al. (2022) from 2–655 days post-merger at microjansky sensitivities, particularly in limiting the parameter space for detectable radio emission from these merger events. Our millijansky sensitivity survey narrows the density parameter space at larger viewing angles due to our observations spanning 501–975 days post-merger.

#### 4.1.2 Events with constraints on jets with $\theta_j$ of $5°$ and $15°$ for $E_{K,iso} \gtrsim 10^{54}$ erg

For the four binary black hole merger events GW170814, GW200129, GW200224, and GW200311, our observational sensitivity could only exclude afterglows for isotropic equivalent jet energies of $\gtrsim 10^{54}$ erg

Constraints on LIGO/Virgo Compact Object Mergers from Late-time Radio Observations    9Table 5. Summary of the constraints on circum-merger density for compact object mergers in all GW events included in the search. These constraints are due to the millijansky sensitivity of the RACS survey, assuming each event launched a relativistic jet viewed off-axis. Both sensitivity and scintillation constraints (see Section 4 for a discussion of both) have been provided at isotropic equivalent jet energies of $2\times10^{51}$, $5\times10^{51}$, $5\times10^{52}$ and $1\times10^{54}$ erg. For each GW event, we provide constraints for tophat and Gaussian jet morphologies at opening angles $\theta_j$ of $5°$ and $15°$. For each considered jet morphology, the first number gives the lower limit for circum-merger densities $n$ that can be excluded given the GW posteriors for inclination angles and the second number gives the viewing angle $\theta_{obs}$ at which this minimum circum-merger density can be ruled out. Refer to Figure 3 and Figures 4-7 in online-only supplementary material.

| | | | Sensitivity Constraints | | | | | | | | Scintillation Constraints | | | | | | | |
|---|---|---|---|---|---|---|---|---|---|---|---|---|---|---|---|---|---|---|
| Jet energies (erg) | | | $2\times10^{51}$ | | $5\times10^{51}$ | | $5\times10^{52}$ | | $1\times10^{54}$ | | $2\times10^{51}$ | | $5\times10^{51}$ | | $5\times10^{52}$ | | $1\times10^{54}$ | |
| GW Name | Jet-Type | $\theta_j$ (°) | n (cm$^{-3}$) | $\theta_{obs}$ (°) | n (cm$^{-3}$) | $\theta_{obs}$ (°) | n (cm$^{-3}$) | $\theta_{obs}$ (°) | n (cm$^{-3}$) | $\theta_{obs}$ (°) | n (cm$^{-3}$) | $\theta_{obs}$ (°) | n (cm$^{-3}$) | $\theta_{obs}$ (°) | n (cm$^{-3}$) | $\theta_{obs}$ (°) | n (cm$^{-3}$) | $\theta_{obs}$ (°) |
| GW170814 | TopHat | 5 | - | - | - | - | - | - | - | - | - | - | - | - | - | - | - | - |
| | | 15 | - | - | - | - | - | - | ≥0.001 | ≤33 | - | - | - | - | - | - | ≥0.001 | ≤28 |
| | Gaussian | 5 | - | - | - | - | - | - | ≥0.09 | ≥48 | - | - | - | - | - | - | ≥1.5 | ≥75 |
| | | 15 | - | - | - | - | - | - | ≥0.001 | ≤42 | - | - | - | - | - | - | ≥0.001 | ≤33 |
| GW170818 | TopHat | 5 | - | - | - | - | - | - | - | - | - | - | - | - | - | - | - | - |
| | | 15 | - | - | - | - | - | - | ≥0.001 | 22 | - | - | - | - | - | - | ≥0.002 | 19 |
| | Gaussian | 5 | - | - | - | - | - | - | - | - | - | - | - | - | - | - | - | - |
| | | 15 | - | - | - | - | - | - | ≥0.001 | ≤30 | - | - | - | - | - | - | ≥0.001 | ≤27 |
| GW190412 | TopHat | 5 | - | - | - | - | - | - | - | - | - | - | - | - | - | - | - | - |
| | | 15 | - | - | ≥3.9 | 27 | ≥7.3 | 27 | ≥0.001 | ≤30 | - | - | ≥5.3 | 27 | ≥7.3 | 27 | ≥0.002 | 27 |
| | Gaussian | 5 | - | - | - | - | ≥3.9 | 27 | ≥2.1 | 27 | - | - | - | - | ≥5.3 | 27 | ≥2.1 | 27 |
| | | 15 | ≥2.8 | 27 | ≥0.6 | 27 | ≥0.05 | 27 | ≥0.001 | ≤36 | ≥7.3 | 27 | ≥1.1 | 27 | ≥0.09 | 27 | ≥0.001 | ≤27 |
| GW190503 | TopHat | 5 | - | - | - | - | ≥7.3 | 10 | - | - | - | - | - | - | ≥8 | 10 | - | - |
| | | 15 | ≥7.3 | 10 | ≥1.1 | 10 | ≥0.02 | 10 | ≥0.001 | 20 | ≥10 | 10 | ≥2.04 | 10 | ≥0.045 | 10 | ≥0.001 | 19 |
| | Gaussian | 5 | - | - | - | - | ≥0.6 | 10 | ≥0.05 | 10 | - | - | - | - | ≥0.8 | 10 | ≥0.06 | 10 |
| | | 15 | ≥1.5 | 10 | ≥0.3 | 10 | ≥0.005 | 10 | ≥0.001 | ≤30 | ≥2.8 | 10 | ≥0.42 | 10 | ≥0.009 | 10 | ≥0.001 | ≤28 |
| GW190701 | TopHat | 5 | - | - | - | - | - | - | - | - | - | - | - | - | - | - | - | - |
| | | 15 | - | - | - | - | - | - | ≥0.001 | 16 | - | - | - | - | - | - | ≥0.009 | 22 |
| | Gaussian | 5 | - | - | - | - | - | - | - | - | - | - | - | - | - | - | - | - |
| | | 15 | - | - | - | - | - | - | ≥0.001 | ≤25 | - | - | - | - | - | - | ≥0.001 | ≤22 |
| GW190814 | TopHat | 5 | - | - | - | - | - | - | 0.007 | 36 | - | - | - | - | - | - | 0.007 | 36 |
| | | 15 | - | - | - | - | - | - | ≥0.001 | ≤39 | - | - | - | - | - | - | ≥0.001 | ≤36 |
| | Gaussian | 5 | - | - | - | - | - | - | ≥0.001 | 36 | - | - | - | - | - | - | 0.005 | 36 |
| | | 15 | - | - | - | - | ≥0.03 | 45 | ≥0.001 | ≤50 | - | - | - | - | ≥0.2 | 62 | ≥0.001 | ≤50 |
| GW200129 | TopHat | 5 | - | - | - | - | - | - | ≥0.09 | 36 | - | - | - | - | - | - | ≥1.08 | 54 |
| | | 15 | - | - | - | - | - | - | ≥0.001 | ≤28 | - | - | - | - | - | - | ≥0.001 | ≤22 |
| | Gaussian | 5 | - | - | - | - | - | - | ≥0.001 | 12 | - | - | - | - | - | - | ≥0.002 | 21 |
| | | 15 | - | - | - | - | - | - | ≥0.001 | ≤36 | - | - | - | - | - | - | ≥0.001 | ≤25 |
| GW200202 | TopHat | 5 | - | - | - | - | - | - | ≥0.001 | ≤24 | - | - | - | - | - | - | ≥0.001 | ≤12 |
| | | 15 | - | - | - | - | - | - | ≥0.001 | ≤33 | - | - | - | - | - | - | ≥0.001 | ≤22 |
| | Gaussian | 5 | - | - | - | - | - | - | ≥0.001 | ≤30 | - | - | - | - | - | - | ≥0.001 | ≤27 |
| | | 15 | - | - | - | - | ≥0.02 | 39 | ≥0.001 | ≤42 | - | - | - | - | ≥0.12 | 47 | ≥0.001 | ≤25 |
| GW200208 | TopHat | 5 | - | - | - | - | - | - | - | - | - | - | - | - | - | - | - | - |
| | | 15 | - | - | - | - | - | - | ≥0.001 | ≤19 | - | - | - | - | - | - | ≥0.001 | ≤19 |
| | Gaussian | 5 | - | - | - | - | - | - | - | - | - | - | - | - | - | - | - | - |
| | | 15 | - | - | - | - | - | - | ≥0.001 | ≤25 | - | - | - | - | - | - | ≥0.001 | ≤25 |
| GW200224 | TopHat | 5 | - | - | - | - | - | - | - | - | - | - | - | - | - | - | - | - |
| | | 15 | - | - | - | - | - | - | ≥0.001 | ≤22 | - | - | - | - | - | - | ≥0.001 | ≤22 |
| | Gaussian | 5 | - | - | - | - | - | - | ≥0.09 | 33 | - | - | - | - | - | - | ≥0.8 | 48 |
| | | 15 | - | - | - | - | - | - | ≥0.001 | ≤27 | - | - | - | - | - | - | ≥0.001 | ≤27 |
| GW200311 | TopHat | 5 | - | - | - | - | - | - | ≥0.3 | 39 | - | - | - | - | - | - | - | - |
| | | 15 | - | - | - | - | - | - | ≥0.001 | ≤25 | - | - | - | - | - | - | ≥0.001 | ≤22 |
| | Gaussian | 5 | - | - | - | - | - | - | ≥0.001 | 18 | - | - | - | - | - | - | ≥0.009 | 24 |
| | | 15 | - | - | - | - | - | - | ≥0.001 | ≤30 | - | - | - | - | - | - | ≥0.001 | ≤25 |

MNRAS 000, 1–17 (2015)

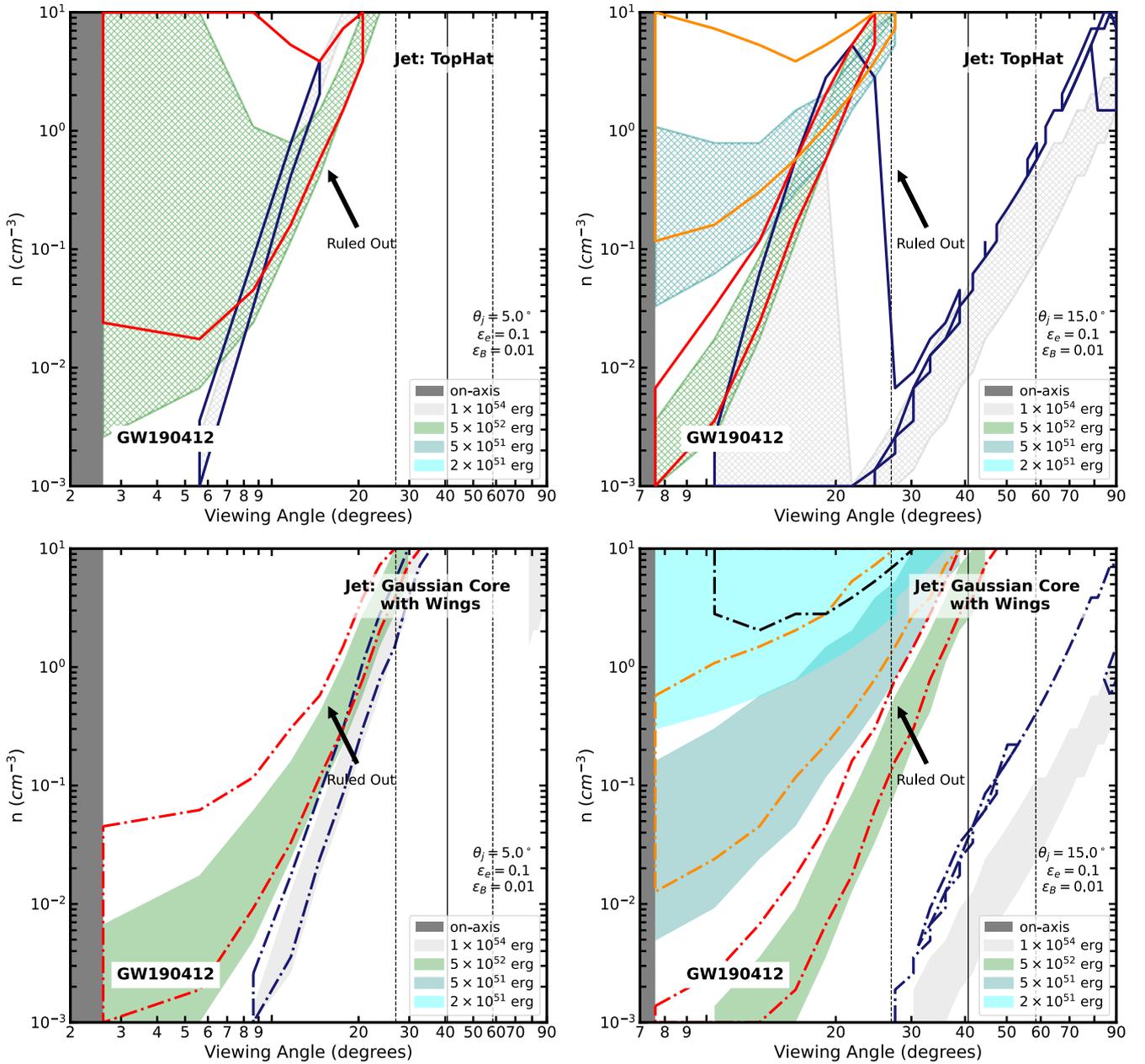

**Figure 3.** Regions of the parameter space ruled out by the non-detection of a counterpart above the $5\sigma$ detection threshold for each of our four epochs at the corresponding days post-merger for the event GW190412. The hatched regions in the top panel show the constraints for the tophat jet with $5°$ and $15°$ opening angles. The filled-in regions in the bottom panel show the constraints for Gaussian jets with $5°$ and $15°$ opening angles and wings spanning up to 4 times the half-opening angle. The colours correspond to the isotropic equivalent kinetic jet energy, $E_{K,iso} = 2 \times 10^{51}, 5 \times 10^{51}, 5 \times 10^{52}$ and $1 \times 10^{54}$ erg. These constraints are a function of microphysical parameters, $p = 2.2$, $\epsilon_e = 0.1$ and $\epsilon_B = 0.01$. The width of each constraint region shows the bounds to the constraints provided by the uncertainty in the distance estimates from the GW detector. Our non-detections exclude all space above and to the left of these limits. The vertical solid and dashed lines indicate the binary inclination angle and associated uncertainty from the GW posteriors. The scintillation constraints are given by the dot-dashed lines for Gaussian jets and by solid lines for tophat jets where the colours - black, dark orange, red and midnight blue correspond to jet energies of $E_{K,iso} = 2 \times 10^{51}, 5 \times 10^{51}, 5 \times 10^{52}$, and $1 \times 10^{54}$ erg, respectively. Additional panels are available in online-only supplementary material.

at all circum-merger densities greater than $0.001$ cm$^{-3}$. This is due to the large distances for these events (see Fig. 4). The lower bounds on $d_L$ are 380, 520, 1070 and 770 Mpc for events GW170814, GW200129, GW200224 and GW200311 respectively while the radio observations start at 630, 358 315 and 292 days post-merger.

Our late-time post-merger observations ranging from 292–1712 days excluded circum-merger densities only at larger viewing angles for 5-degree jet opening angles (see Figures 4-7 in online-only sup-plementary material). This is because the lightcurve for narrower jets may have already decayed at 300 days post-merger as compared to a wider one, hence, potential emission, especially from narrow or low-energy jets, may be too dim to be constrained (see Fig. 5). For GW170814 and GW200224, we cannot rule out circum-merger densities for smaller viewing angles at the upper distance limit from the GW data. For GW200311, we cannot constrain tophat jets with 5-

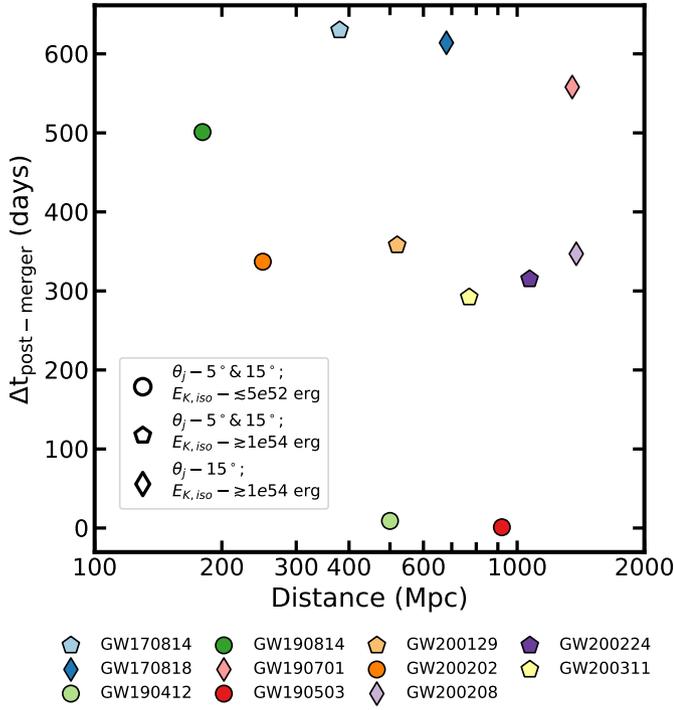

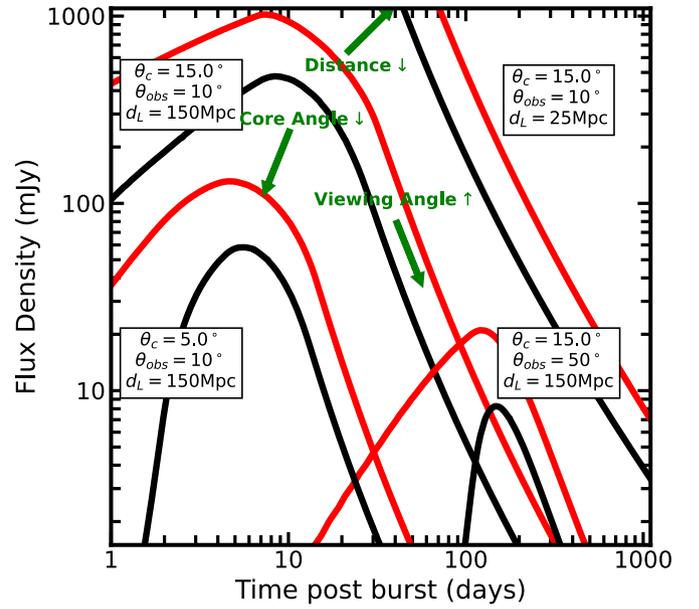

Figure 4. Summary of constraints for events included in our search as a function of days post-merger that the observations started and the lower bound (5th percentile) on the luminosity distance from the GW data. The shapes denote the range of viewing angles and energies where our observations can provide constraints.

degree opening angles as the detectable parameter space lies outside the viewing angles allowed by the GW posterior.

### 4.1.3 Events with constraints on jets with $\theta_j$ of 15° for $E_{K,iso} \gtrsim 10^{54}$ erg

For the binary black hole merger events GW170818 and GW190701 (and GW200208, if the candidate is not associated with the event), our observational sensitivity allowed the exclusion of circum-merger densities of $n \geq 0.001$ cm$^{-3}$ only for jets with wider opening angles of 15° at high isotropic equivalent jet energies of $\gtrsim 10^{54}$ erg. This is due to the large distances for these events (lower bound on d$_L$ $\geq$680 Mpc) and post-merger observations at $\geq$558 days (see Fig. 4). For the upper bound of distance limits, we cannot exclude lower inclination angles because the lightcurve may have already decayed beyond the detection threshold for low inclination angles (see Fig. 5). The constraints highlight the importance of prompt post-merger monitoring to better constrain detectable counterparts.

### 4.2 Luminosity Constraints

With the distance localisation provided by the Advanced LIGO/Virgo detectors, we can calculate the equivalent luminosity at 888 MHz for our 5-sigma survey limit of 1.5 mJy (see Table 6). This luminosity provides an upper limit on the accompanied afterglow in case of no detections, considering that the source lies within its 90% credibility region, using $L_{rad} = 4\pi d_L^2$, where we use the median value for the distance luminosity estimates from the GW analysis. These limits ranged from $8.5 \times 10^{37}$ erg s$^{-1}$ to $8.0 \times 10^{39}$ erg s$^{-1}$ at 888 MHz.

Figure 5. Temporal evolution of off-axis lightcurves for tophat (black) and Gaussian (red) relativistic jets as a function of core angle, viewing angle, and event distance. The core angle $\theta_c$, viewing angle $\theta_{obs}$ and event distance $d_L$ for each pair of lightcurves are given in the adjacent box. All lightcurves are computed for a jet $E_{K,iso}$ of $5 \times 10^{52}$ erg.

Table 6. Luminosity limits on the radio counterparts of GW events in our search, based on our 5-sigma survey sensitivity limit of 1.5 mJy. Column 1 gives the gravitational-wave event and Column 2 gives the upper limit on the radio luminosity for a region with no potential candidates.

| GW Event | $L_{radio, 888MHz}$ (erg s$^{-1}$) |
| --- | --- |
| GW170814 | $5.8 \times 10^{38}$ |
| GW170818 | $1.8 \times 10^{39}$ |
| GW190412 | $8.4 \times 10^{38}$ |
| GW190503 | $3.7 \times 10^{39}$ |
| GW190701 | $7.1 \times 10^{39}$ |
| GW190814 | $8.5 \times 10^{37}$ |
| GW200129 | $1.3 \times 10^{39}$ |
| GW200202 | $2.7 \times 10^{38}$ |
| GW200208 | $8.0 \times 10^{39}$ |
| GW200224 | $4.7 \times 10^{39}$ |
| GW200311 | $2.2 \times 10^{39}$ |

### 4.3 Synchrotron Self-Absorption Effects on Constraints

Synchrotron self-absorption (SSA) becomes important at lower frequencies as low-energy synchrotron photons are reabsorbed by the same relativistic electrons that emitted them (Condon & Ransom 2016). SSA suppresses emission at lower frequencies, causing a spectral turnover where the spectral index is strongly inverted, with $\alpha \geq 1$ (Granot & Sari 2002; Mészáros 2006). To address the effect of SSA on our results, we use the code PyBlastAfterglowMag[10] which compares well with afterglowpy results under non-SSA conditions (Nedora et al. 2024).

We used PyBlastAfterglowMag to simulate a grid of light curves with self-absorption effects between 0.888 and 1.65 GHz for two relativistic jets: a tophat jet with a jet opening angle, $\theta_j$, of 5° and 15°; and a Gaussian jet with a jet core $\theta_j$ of 5° and 15° and wings extending to four times the jet half opening angle i.e. 10° and 30° respectively. We used a Lorentz factor of 200 for a tophat jet simulation and 800 for a Gaussian jet simulation. This adjustment was necessary due to numerical simulation limitations when using smaller Lorentz factors for Gaussian jets. We generated these grids at isotropic-equivalent kinetic energies, $E_{K,iso}$, of $2 \times 10^{51}$ and $1 \times 10^{54}$ erg at an event distance of 180 Mpc. We computed the light curves for off-axis viewing angles ranging from 2.6° – 90° and 7.6° – 90° for jets with 2.5° and 7.5° half-opening angles respectively and fixed circum-merger densities from $10^{-3}$ to 10 cm$^{-3}$. We extracted the maximum number of days post-burst when the afterglow is detectable (flux density > 1.5 mJy) and inverted ($\alpha \geq 0$) in the synchrotron self-absorption regime, as per the search criteria applied to our dataset in Section 2.2. Fig. 6 shows the dependence of the maximum time (in days) that the afterglow is observed as inverted in the SSA regime on the viewing angle and ambient density. To assess the impact of SSA on our constraints, we identify the latest post-burst time at which the afterglow remains both detectable (flux density > 1.5 mJy) and inverted ($\alpha \geq 1$) solely due to SSA. The omission of SSA in `afterglowpy` models could have led to an overestimation of flux at our frequencies, potentially allowing for tighter constraints for a region of our parameter space. Fig. 6 highlights the 300-day limit for SSA-driven inversion as a function of the viewing angle and density for $E_{K,iso} = 1 \times 10^{54}$ erg.

---

[10] https://github.com/vsevolodnedora/PyBlastAfterglowMag

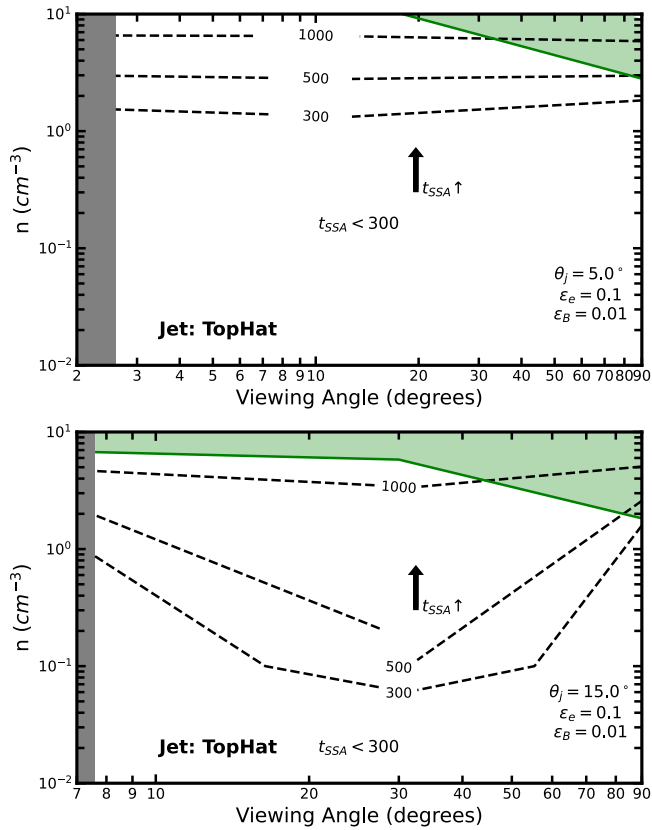

**Figure 6.** The maximum number of days that an afterglow may be inverted in a self-absorption regime for a merger event as a function of circum-merger density and viewing angle. The contours mark the regions of parameter space above which an afterglow may be self-absorbed for 300, 500 and 1000 days as a function of circum-merger density and viewing angle for isotropic equivalent jet energy $E_{K,iso} \in \{1 \times 10^{49}, 1 \times 10^{54}\}$ erg. The dashed lines correspond to $1 \times 10^{54}$ erg respectively. The top panels show the effect of SSA for off-axis afterglows of tophat jets with $\theta_j$ of 5° and the bottom panels for tophat jets with $\theta_j$ of 15°. The grey-shaded region corresponds to on-axis afterglows. Additional panels for Gaussian jets are available in online-only supplementary material. The green-shaded region represents the area where the afterglow is inverted beyond 300 days solely due to SSA ($\alpha \geq 1$) for jets with an $E_{K,iso}$, of $1 \times 10^{54}$ erg.

In our frequency range, synchrotron self-absorption (SSA) effects typically last for a short period and rarely persist beyond 300 days. For lower-energy jets ($2 \times 10^{51}$ erg), the afterglow spectra are no longer inverted and detectable at late times, regardless of density or viewing angle. Even for moderate densities ($\gtrsim 0.1$ cm$^{-3}$), SSA effects fade well before our typical post-merger observation timespan. Only in dense environments and for high-energy jets ($1 \times 10^{54}$ erg) do SSA effects persist for longer durations (> 300 days). For larger core angles, the jet carries greater total energy, leading to stronger self-absorption even at lower circum-merger densities of $\gtrsim 0.1$ cm$^{-3}$. These findings are consistent with the expected behaviour of the synchrotron self-absorption frequency, which scales directly with jet energy, circum-merger density, and $\epsilon_B$, and inversely with time post-merger (Granot & Sari 2002). The maximum time an afterglow is inverted appears to be largely independent of the jet structure (see online-only supplementary material). The only notable difference arises from the use of a higher Lorentz factor for Gaussian jets, which leads to delayed emission.

Using this information, we can ascertain the likelihood that a true afterglow may be excluded based on our spectral index cut in (vi) of 2.2. In our search, we excluded 32 candidates with inverted spectra in the GW190503 region, 7 candidates in the GW170814 region, and 3 candidates in the GW170818 region. For GW190503, although the first observation occurred only one day post-merger, it is unlikely that a genuine afterglow was excluded based on having a constant inverted spectrum. An off-axis afterglow would have had negligible flux at day one, making a constant spectrum fit impossible, while an on-axis afterglow would rapidly evolve, quickly becoming optically thin and thus not presenting a constant inverted spectrum. For GW170814 and GW170818, the first observations were made ~ 600 days after the merger, which means that we could have only missed an afterglow if it was in a dense environment. Hence, our decision to remove inverted spectrum sources will miss only a small fraction of potential afterglows that may be inverted in the SSA regime. Furthermore, our SSA analysis suggests that for jets with higher energies (1 × $10^{54}$ erg), we cannot effectively constrain afterglows that may be in dense environments ($\geq 1$ cm$^{-3}$), especially for larger viewing angles. At a given jet energy and circum-merger density, a larger viewing angle leads to SSA effects remaining relevant for longer (in the observer frame), although few such large-viewing-angle events would be detectable in any case.

## 5 ASKAP DETECTABILITY AND RATES

In this section, we discuss the prospects and usefulness of future late-time radio searches for compact object merger afterglows with millijansky sensitivity surveys.

### 5.1 ASKAP Detectability

We identified potential afterglow candidates within the 90% posterior probability region for three out of the 11 events we investigated. However, none of these candidates are likely to be associated with the corresponding GW event. Four out of the 11 events discussed in Section 4, are well constrained for both wide and narrow opening angles and typical afterglow energies ($\lesssim 5 \times 10^{52}$ erg); four are only constrained at high isotropic energies for both opening angles considered and three are only constrained at high isotropic energies ($1 \times 10^{54}$ erg) and wider opening angles. For fixed isotropic equivalent energy, narrower jets are less energetic in terms of total energy, and thus easier to miss, so constraints are weaker as seen in in Section 4. This provides a rationale for investigating the usefulness of late-time radio searches for afterglows with millijansky-sensitivity surveys, to determine if we can place useful constraints or detect an afterglow for a compact object merger at a certain distance given our observational cadence in relation to the merger.

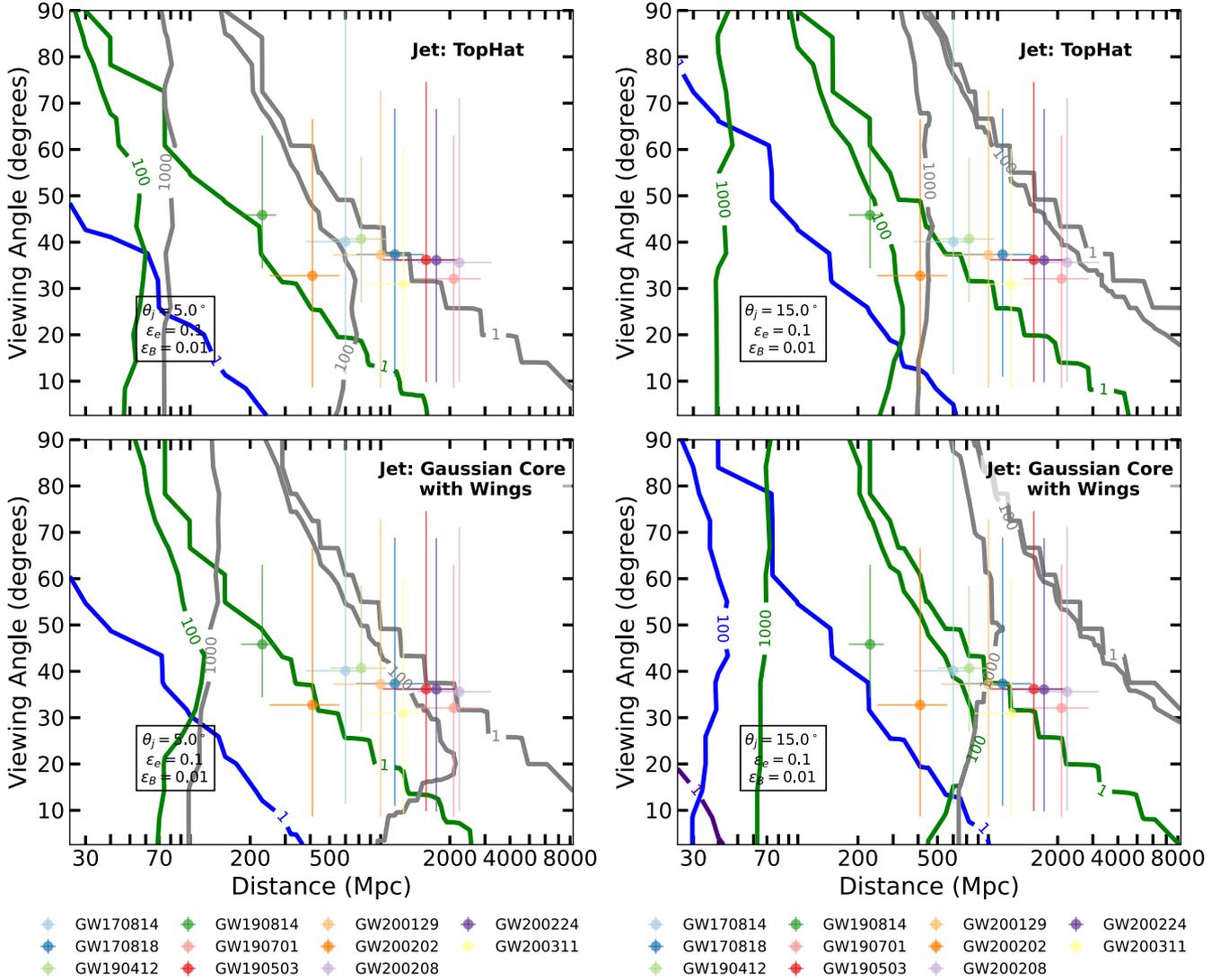

**Figure 7.** The number of days that an afterglow is detectable for a merger event with a survey at a sensitivity of 1.5 mJy (5$\sigma$) as a function of event distance and viewing angle. The contours mark the regions of parameter space above which an afterglow will be detectable for 1, 100 and 1000 days by a millijansky level survey as a function of distance and viewing angle for isotropic equivalent jet energy $E_{K,iso} \in \{1 \times 10^{49}, 1 \times 10^{54}\}$ erg. Contour colors indigo, blue, green and grey correspond to $E_{K,iso}$, of $1 \times 10^{49}$, $2 \times 10^{51}$, $5 \times 10^{52}$ and $1 \times 10^{54}$ erg respectively. The top panels show the detectability for off-axis afterglows of tophat jets with $\theta_j$ of 5° (left) and 15° (right) and the bottom panels correspond to off-axis afterglows for Gaussian jets with $\theta_j$ of 5° (left) and 15° (right) and wings that span up to 10° and 30°, respectively, on each side. The coloured markers represent the 11 events included in our search; for these, the uncertainty in the ordinate and abscissa span the 90% credible interval on the inclination angle and merger distance GW posteriors respectively.

The constraints discussed are intricately tied to the viewing angle, distance, jet structure, and our observational sampling post-merger.

To address this question of detectability of afterglows with a millijansky sensitivity survey, we compute the limit number of days an afterglow is detectable at distances ranging from 25 to 8000 Mpc. We use `afterglowpy` to generate a grid of light curves at 888 MHz for two relativistic jets: a tophat jet with a jet opening angle, $\theta_j$, of 5° and 15°; and a Gaussian jet with a jet core $\theta_j$ of 5° and 15° and wings extending to 4 times the jet half opening angle i.e. 10° and 30°. Given the uncertainty in the observational evidence for the environments of compact mergers (see for eg., Tunnicliffe et al. 2014; Fong et al. 2015), we assume these events occur in a uniform interstellar medium (ISM) with a density of 1 cm$^{-3}$ (Tak et al. 2019; Santana et al. 2014). We assume the microphysical parameters $p = 2.2$, $\epsilon_e = 0.1$, and $\epsilon_B = 0.01$, as discussed previously. We generate these grids at isotropic-equivalent kinetic energies, $E_{K,iso}$, of $1 \times 10^{49}$, $2 \times 10^{51}$, $5 \times 10^{52}$ and $1 \times 10^{54}$ erg. For each isotropic en-ergy value, we extract a series of contour plots for off-axis afterglow that are detectable above 1, 100 and 1000 days above ∼ 1.5 mJy. The contours illustrate regions of detectability as a function of distance and viewing angle for different energy levels, which can be used to understand how far and at what angles an off-axis afterglow can be detected, given a certain jet energy (see Fig. 7). For example, for a future event, with an estimated viewing angle of ∼ 10° and a distance of ∼ 1500 Mpc, we can constrain afterglow emission for a jet down to an energy of $5 \times 10^{52}$ erg with a survey with epoch of separation less than a day.

Using the detectability diagnostic, we find that (i) off-axis afterglow detectability for tophat jets falls off more steeply with viewing angle as compared to structured Gaussian jets, and (ii) wider jet cores with the same isotropic-equivalent energy have more total energy than a narrow jet and, hence, are easier to detect at large viewing angles and distances. In Fig. 7, we compare the estimated distances and viewing angles derived from GW data for our sample of events with the expected detectability of afterglows. As anticipated, events with assumed higher energy wide-angle jets in conjunction are more likely to be detectable, which helps us constrain jet properties. For events at larger distances, observing their off-axis afterglows within 100 days post-merger, as in the cases of GW190412 and GW190503, allows us to place tighter constraints on jet properties.

### 5.2 Rates

Here, we address the issue of how many afterglow-like transients we can expect for future localisation. This has a direct impact on the feasibility of the search for late-time afterglows with a survey of this depth and cadence and the inferred credibility of identified candidates. In this subsection, we discuss a general rate for afterglow-like transients we can expect for a future GW event with a millijansky-sensitivity survey.

In our main search, we have searched a total of ∼600 deg², i.e. ≲2% of the ∼36,000 deg² RACS footprint. To provide estimates on the number of afterglow-like transients for a future gravitational wave event, we use all radio sources in the RACS footprint and apply the automated criteria used for our search in Section 2. We used a random sampling of 50 ellipses, each with an area of 1 deg², within the RACS footprint to report the average number of radio sources in RACS per square degree to be 80±23. We then compare the modulation index for RACS sources in each of these regions to the $m_{RISS}$ ranging from 0.02-0.33, to extract the number of variable sources in each random region. We use this range of $m_{RISS}$ to calculate the number of variable sources that can be expected for a future GW event localisation at any Galactic latitude. The top panel of Fig. 8 shows the average number of variable RACS sources over these randomly sampled regions for each $m_{RISS}$ cut. As the threshold for artefact variability from extragalactic sources, $m_{RISS}$, increases, a larger number of variable sources become consistent with scintillation.

Not all variable sources in a given region will match the expected afterglow flux density, as this depends heavily on the event's distance and the energy of the jet. For a given jet energy and event distance, we calculate the maximum possible off-axis afterglow flux density. This maximum occurs at the smallest inclination angle of 7.6°, for an assumed wide jet with a half-opening angle of 7.5°. We compute the flux density for tophat and Gaussian jets at isotropic equivalent energies $E_{K,iso}$ ranging from $1 \times 10^{49}$ to $1 \times 10^{54}$ ergs, over distances from 25 to 8000 Mpc using afterglowpy. We assume these events occur in a uniform interstellar medium (ISM) with a density of 1 cm$^{-3}$, as discussed previously. We assume the microphysical parameters $p = 2.2$, $\epsilon_e = 0.1$, and $\epsilon_B = 0.01$, as discussed previ-ously. We then filter out sources with flux densities higher than the maximum expected for an off-axis afterglow with an assumed wide opening angle of 7.5°, a correspondingly low viewing angle of 7.6° for the variable sample with the lowest expected modulation index of 0.02. The remaining sources are expressed as a percentage of the total variable sources in that region. The bottom panel of Fig. 8 illustrates the average percentage (over 50 randomly sampled regions) of variable sources consistent with a viable afterglow flux density, plotted as a function of event distance and jet energy.

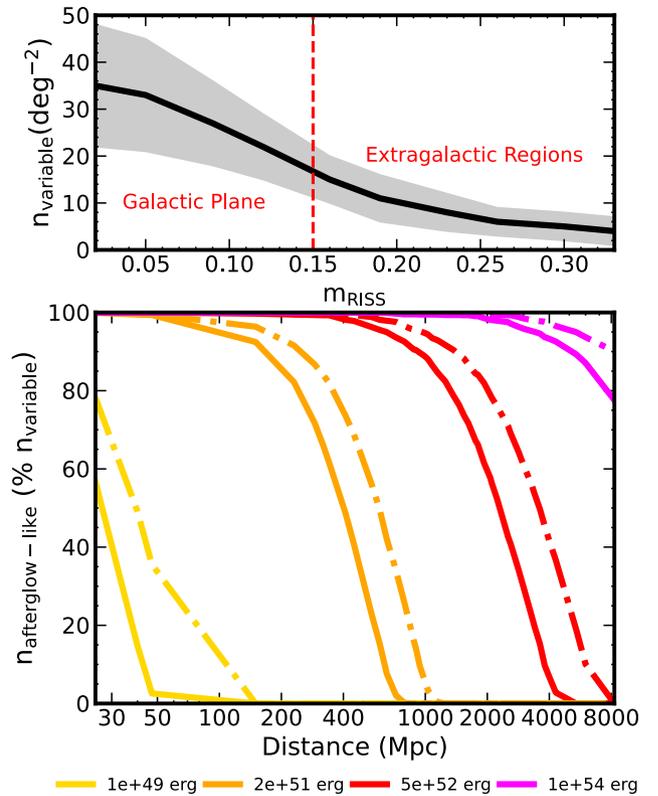

**Figure 8.** Average number of variable and afterglow-like transients expected from a future GW localisation. The top panel shows the average number of variable sources not consistent with scintillation per square degree for 50 randomly centred 1 square-degree ellipses on the sky within the RACS footprint. The shaded region gives the $1\sigma$ spread. The dashed line marks the modulation index distinction between the regions in the Galactic plane and outside it. The bottom panel gives the fraction of afterglow-like transients as an average percentage of variable sources within the expected flux density of an off-axis afterglow. We assumed the off-axis afterglow for a relativistic jet launched with a half opening angle of 7.5°, isotropic equivalent energy from $E_{K,iso} \in \{1 \times 10^{49}, 1 \times 10^{54}\}$ erg at distances in the range 25-8000 Mpc, being viewed at an angle of 7.6°. The solid dashed and the dotted dashed lines correspond to tophat and Gaussian jets respectively.

We can use the results shown in Figure 8 to calculate the number of off-axis late-time afterglow-like transients in millijansky-level surveys for a future compact object merger localisation. For example, consider a compact object merger with a 90% posterior localisation of 20 deg² on the sky and the average modulation index due to RISS of the corresponding region of 0.25. Figure 1 shows the contours that can be used to estimate the modulation index for localisation on the sky at 888 MHz. The number of variable sources above the scintillation level of 0.25 is 6 ± 2 per square degree from the top panel of Fig. 8. Assuming a merger distance of 300 Mpc, we use the bottom panel of Fig. 8 to assess the percentage of variable sources that will be within the expected maximum afterglow flux density for each jet energy. Assuming an afterglow for a tophat jet with $E_{K,iso}$ of $1 \times 10^{49}, 2 \times 10^{51}, 5 \times 10^{52}$ and $1 \times 10^{54}$ erg, we would expect $\sim 0\%, 70\%, 100\%$, and $100\%$ of the 6 variable sources per square degree to be within the maximum flux density expected for an off-axis afterglow at that distance. This gives us 0, 5, 6, and 6 variables consistent with an afterglow-like transient. Multiplying by the entire localisation area of 20 deg$^2$ yields $\sim$ 0, 100, 120, and 120 variable sources that are consistent with an off-axis afterglow flux density for the chosen $E_{K,iso}$ in this region.

It should be noted that this sample would still be contaminated by other variable radio sources and individual estimates have not been provided for each of the filtering criteria in Section 2. However, rough estimates from our small sample suggest (i) $\geq 80\%$ of variable sources were identified and eliminated through archival radio detections in the northern hemisphere, (ii) $\geq 47\%$ were successfully identified and excluded for all southern events (declination $\leq -40$ deg). For the sources not filtered out in the archival radio observations cut in the southern hemisphere events, an additional $\sim 34\%$ were eliminated through additional ASKAP data points in which the variable sources were persistent and $\sim 22\%$ were removed by fitting their spectrum against RACS mid-high and low observations. Hence, archival observations are informative for ruling out potential candidates in future runs. The cut designed to filter out sources with radio emission pre-burst may also remove potential host galaxies with radio emission. An afterglow may occur in an already radio-bright galaxy, which may have been removed according to the aforementioned cut. To calculate the rate of false negatives, i.e. potential host galaxies eliminated in this cut, we cross-match the eliminated sources with galaxies in the GLADE+ catalogue. We use offsets of $\leq 30$ kpc (Fong et al. 2010) within the merger volume (propagated to astrometric offsets using the distance estimates from the GW detectors). We find an average of 5.5% of eliminated sources would have a match to a potential galaxy, i.e., would lie within the expected distance and physical offsets, in GLADE+.

# 6 DISCUSSION AND CONCLUSION

In our work, we find potential afterglow candidates for BBH events – GW190503, GW200202 and GW200208. On further inspection we find that the candidates for events GW190503 and GW200202 can be conclusively ruled out as unrelated to the event. For the candidate in GW200208's localisation, while it does not fit existing afterglow models well, its radio luminosity of $10^{40}$ erg s$^{-1}$ and galaxy offsets of 22-54 kpc are within ranges typical of known radio afterglows (Chandra & Frail 2012; Fong et al. 2015; Mooley et al. 2016). However, we note that the average $m_{RISS}$ for this localisation is quite low, implying a high false positive rate for the identification of a source as a candidate, hence, it is unlikely to be related to the gravitational-wave event.

Since, we found no likely candidates in our search, we examine the observational constraints that radio observations can place on the circum-merger environment and jet energetics of relativistic jets that may be launched by compact object mergers. We can impose stringent constraints for four out of 11 events observed at closer distances (lower bound on $d_L \leq 250$ Mpc) or within a few days post-merger. We can constrain only higher energy jets for four that were observed at distances $\leq 600$ Mpc (lower bound of GW distance posterior) or observed within $\sim 300$ days post-merger. We cannot constrain lower energy or narrow jets for three that were observed at distances $\geq 600$ Mpc (lower bound of GW distance posterior) or observed at $\sim$ 600 days post-merger. Our constraints are not sensitive to synchrotron self-absorbed afterglows for higher jet energies ($E_{K,iso} \sim 10^{54}$ erg) in dense circum-merger environments and highly off-axis viewing angles.

Most of the known radio-detected short-GRBs are found in environments with a circum-merger density of $< 1$ cm$^{-3}$ (Schroeder et al. 2024). The density constraints derived from our analysis are not sensitve enough to probe lower jet energies ($E_{K,iso} \leq 10^{54}$ erg). These weak constraints arise from ASKAP's sensitivity (1.5 mJy), scintillation effects, and limited temporal coverage. The sensitivity restricts detection of faint afterglows from low-energy jets or distant, narrow cores. Sparse observation cadence further reduces the likelihood of capturing afterglows at peak visibility, especially for specific viewing angles. High refractive interstellar scintillation (RISS) thresholds combined with low cadence complicate identifying true afterglows. Improved survey sensitivity and cadence, such as collaborative efforts with VAST's bi-monthly surveys, will enhance detectability, filter artefacts, and support a RISS-agnostic approach to identifying afterglows.

Our analysis demonstrates that millijansky sensitivity surveys can detect radio afterglows at late times, making late-time searches for EM counterparts feasible. However, several limitations persist. We assume jetted emission from compact object mergers. While the presented observations set new sensitivity limits for well-localised BBH mergers in the O2 and O3 runs, current radio constraints remain weak due to the timing of observations and their sensitivity in relation to the merger. Additionally, radio observations are ineffective for detecting EM signatures in galaxies with radio emission or bright AGN. Late-time searches may also struggle to identify afterglows that have become extended and may no longer appear variable. This is because compact sources, which vary more rapidly, are easier to detect above scintillation levels, while extended afterglows, varying more slowly, may fall below the variability threshold (Dobie et al. 2020).

Since we do not expect EM emission from BBH mergers and follow-up is not feasible for huge localisations, BBHs are not prime targets for follow-up. So despite these weak constraints, using the widefield radio surveying capability of telescopes such as ASKAP is the only strategy that enables observational constraints on afterglows of the potential relativistic jets that may be launched by compact object mergers. Unlike other multiwavelength follow-ups, where optical or X-ray observations (Yang et al. 2019; Gompertz et al. 2020; Kim et al. 2021) may be dominated by isotropic radiation from nuclear-decay-powered kilonovae shortly after the merger (within 2-3 days), late-time radio observations offer the opportunity to constrain the environments of compact object mergers if a jet was launched by probing the synchrotron radiation from a jet interacting with the CSM. This is because radio counterparts persist longer than other wavelengths, allowing constraints to be placed at late times, over hundreds of days post-merger.

# ACKNOWLEDGEMENTS

We would like to acknowledge the anonymous reviewer whose suggestions have helped to improve this manuscript. We would also like to acknowledge Alexander van der Horst for his advice about SSA effects. DLK was supported by NSF grant AST-1816492. This research was conducted with support from the Australian Research Council Centre of Excellence for Gravitational Wave Discovery (OzGrav), project number CE230100016. This scientific work uses data obtained from Inyarrimanha Ilgari Bundara / the CSIRO's Murchison

Radio-astronomy Observatory. We acknowledge the Wajarri Yamaji People as the Traditional Owners and native title holders of the Observatory site. CSIRO's ASKAP radio telescope is part of the Australia Telescope National Facility. Operation of ASKAP is funded by the Australian Government with support from the National Collaborative Research Infrastructure Strategy. ASKAP uses the resources of the Pawsey Supercomputing Research Centre. Establishment of ASKAP, Inyarrimanha Ilgari Bundara, the CSIRO Murchison Radio-astronomy Observatory and the Pawsey Supercomputing Research Centre are initiatives of the Australian Government, with support from the Government of Western Australia and the Science and Industry Endowment Fund.

## DATA AVAILABILITY

The ASKAP data used in this work is available from the CSIRO ASKAP Science Data Archive, CASDA[11] under the project codes AS107, AS110 and AS207.

---

[11] http://data.csiro.au/

This paper has been typeset from a T<sub>E</sub>X/L<sup>A</sup>T<sub>E</sub>X file prepared by the author.

# Supplementary Material for "Constraints on LIGO/Virgo Compact Object Mergers from Late-time Radio Observations"

Ashna Gulati,[1,2,3]* Tara Murphy,[1,2] Dougal Dobie,[1,2,5] Adam Deller,[2,5] David L. Kaplan,[7] Emil Lenc,[3] Ilya Mandel,[6,2] Stefan Duchesne,[4] Vanessa Moss[3]

[1]*Sydney Institute for Astronomy, School of Physics, The University of Sydney, NSW 2006, Australia*
[2]*ARC Centre of Excellence for Gravitational Wave Discovery (OzGrav), Hawthorn, VIC 3122, Australia*
[3]*CSIRO Space and Astronomy, PO Box 76, Epping, NSW 1710, Australia*
[4]*CSIRO Space and Astronomy, PO Box 1130, Bentley WA 6102, Australia*
[5]*Centre for Astrophysics and Supercomputing, Swinburne University of Technology, Hawthorn, Victoria, Australia*
[6]*School of Physics and Astronomy, Monash University, Clayton, Victoria 3800, Australia*
[7]*Department of Physics, University of Wisconsin-Milwaukee, P.O. Box 413, Milwaukee, WI 53201, USA*

19 March 2025

## 1 ADDITIONAL FIGURES

Here we provide additional figures referenced in the main text.

In Fig. 1, we present the WISE color-color diagram for the candidate afterglow in the localisation of GW200208 as discussed in Section 3 of the main paper. In Fig 2, we present afterglow model fits to the unlikely candidates in the localisation regions for GW190503 and GW200202. In Fig 3, we present the cornerplot for the MCMC sampling for a TopHat afterglow fitting to the candidate synchrotron source in the GW200208 sky localisation.

In Figures 4–7, we include the constraint curves for all GW events other than GW190412 which has already been presented and discussed in Section 4 of the main paper. In Fig 8, we include the SSA effects on off-axis afterglows for Gausian jets for the same core angles that have been presented and discussed in Section 4 of the main paper.

## REFERENCES

Wright E. L., et al., 2010, AJ, 140, 1868

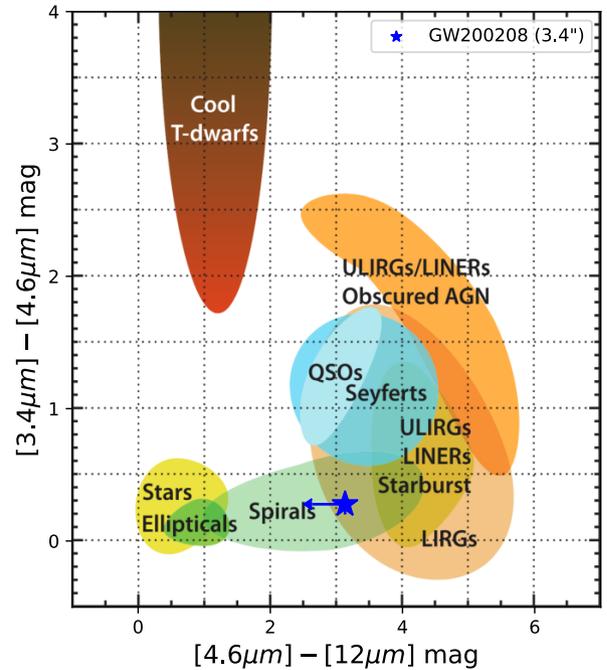

**Figure 1.** Colour-colour diagram of the WISE Infrared source associated with the candidate for event GW200208 discussed in Section 3. The background colours give the classification scheme from Wright et al. (2010). However, any classification from WISE colours alone is limited as there are no detections in the 12$\mu$m band.

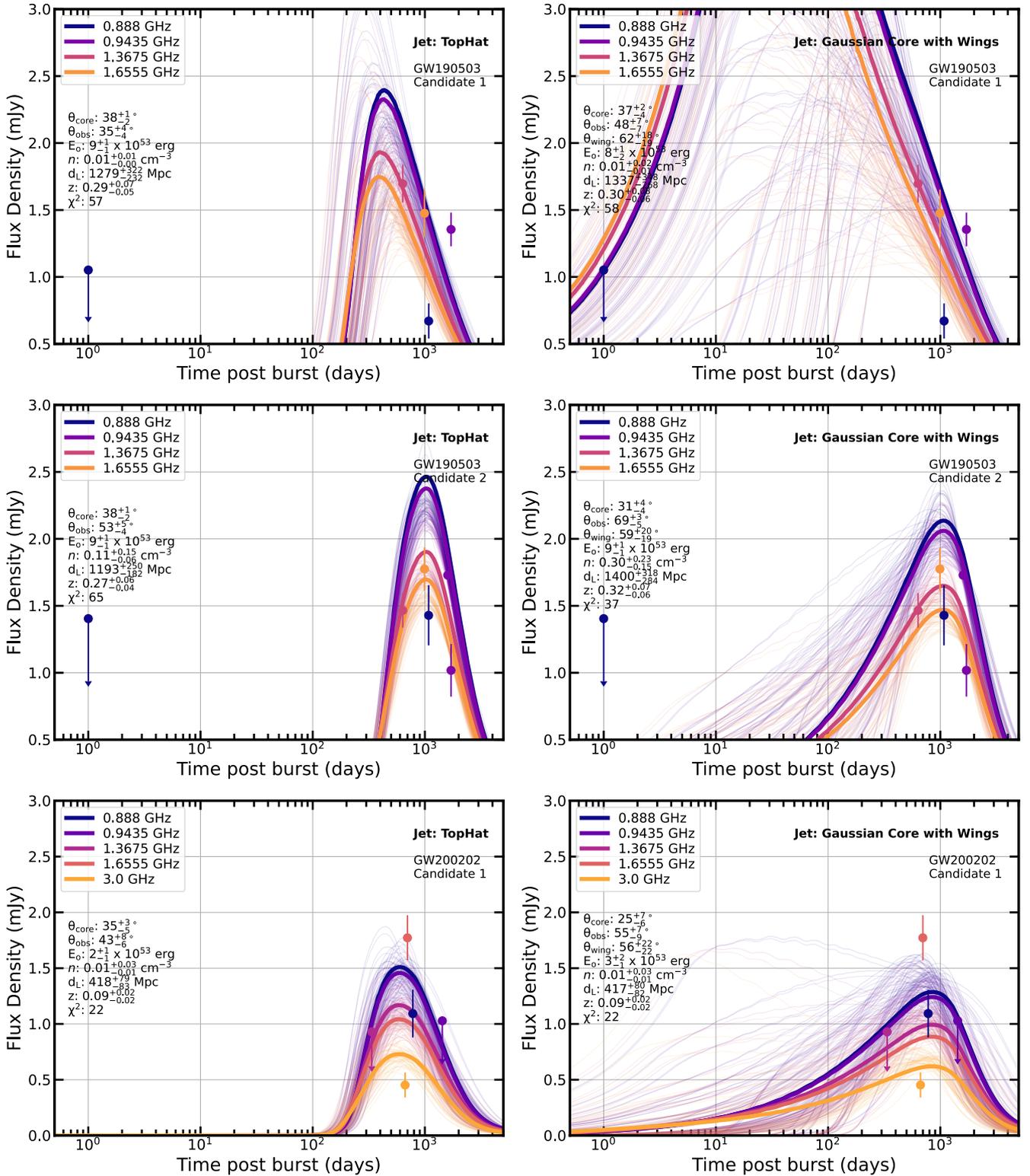

**Figure 2.** TopHat and Gaussian jet models fit to potential candidates. Other details in Fig. 2 of the main paper.

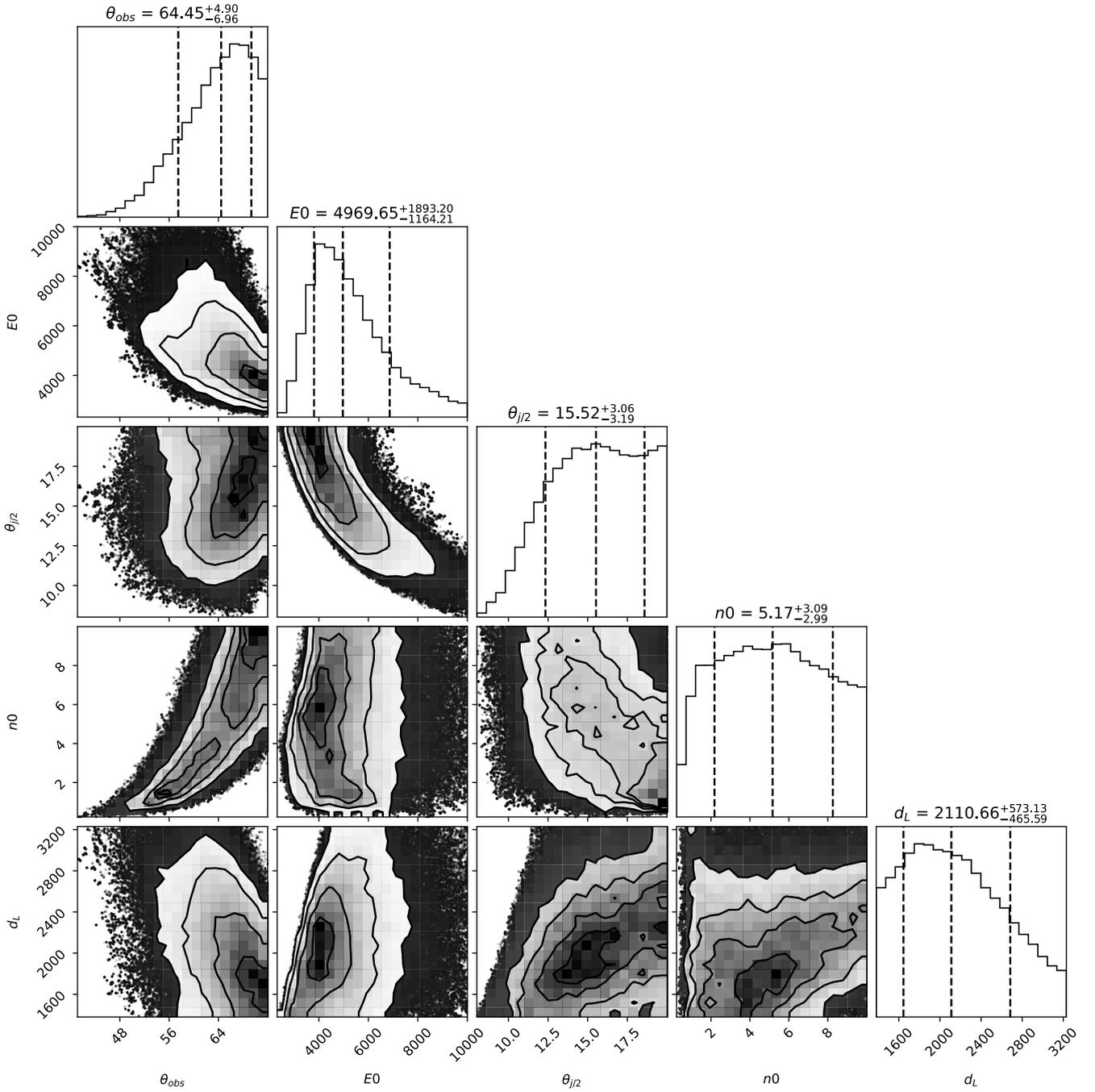

**Figure 3.** Cornerplot for the MCMC TopHat fit for the candidate synchrotron source at the 79.5% posterior percentile of the GW200208 sky localisation.

# Events with Constraints on jets with $\theta_j - 5°$ and $15°$ for $E_{K,iso} \lesssim 5 \times 10^{52}$ erg

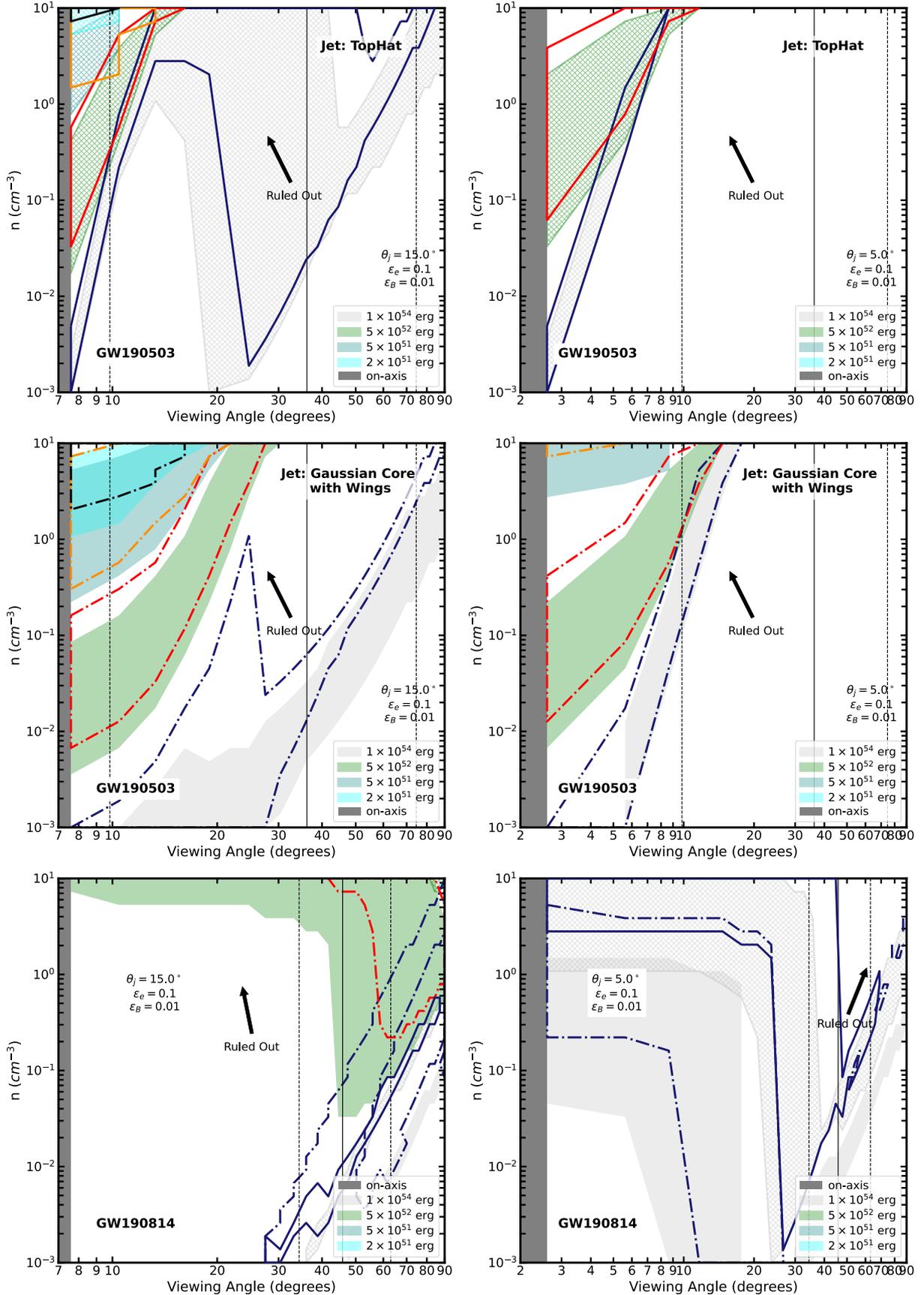

**Figure 4.** Observational constraints for events GW190503 and GW190814. Other details in Fig. 3 of the main paper.

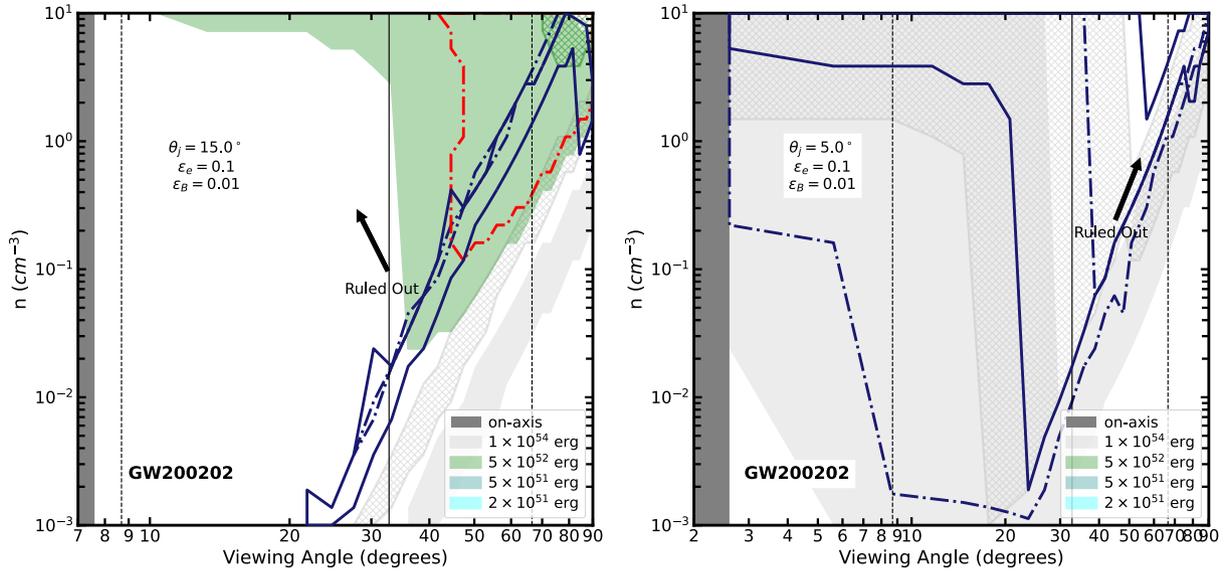

**Events with Constraints on jets with $\theta_j - 5°$ and $15°$ for $E_{K,iso} \gtrsim 10^{54}$ erg**

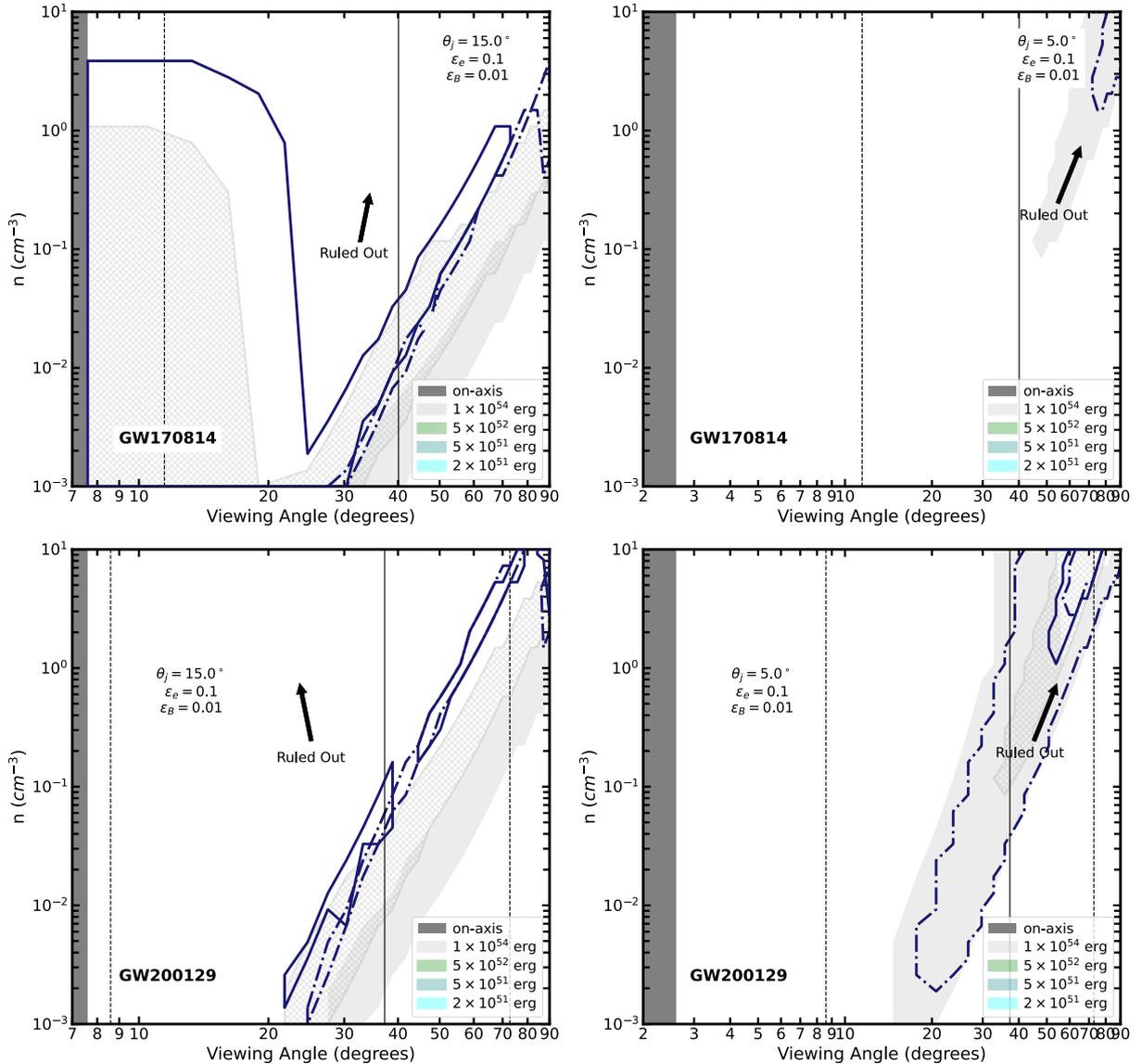

**Figure 5.** Observational constraints for events GW200202, GW170814 and GW200129. Other details in Fig. 3 of the main paper.

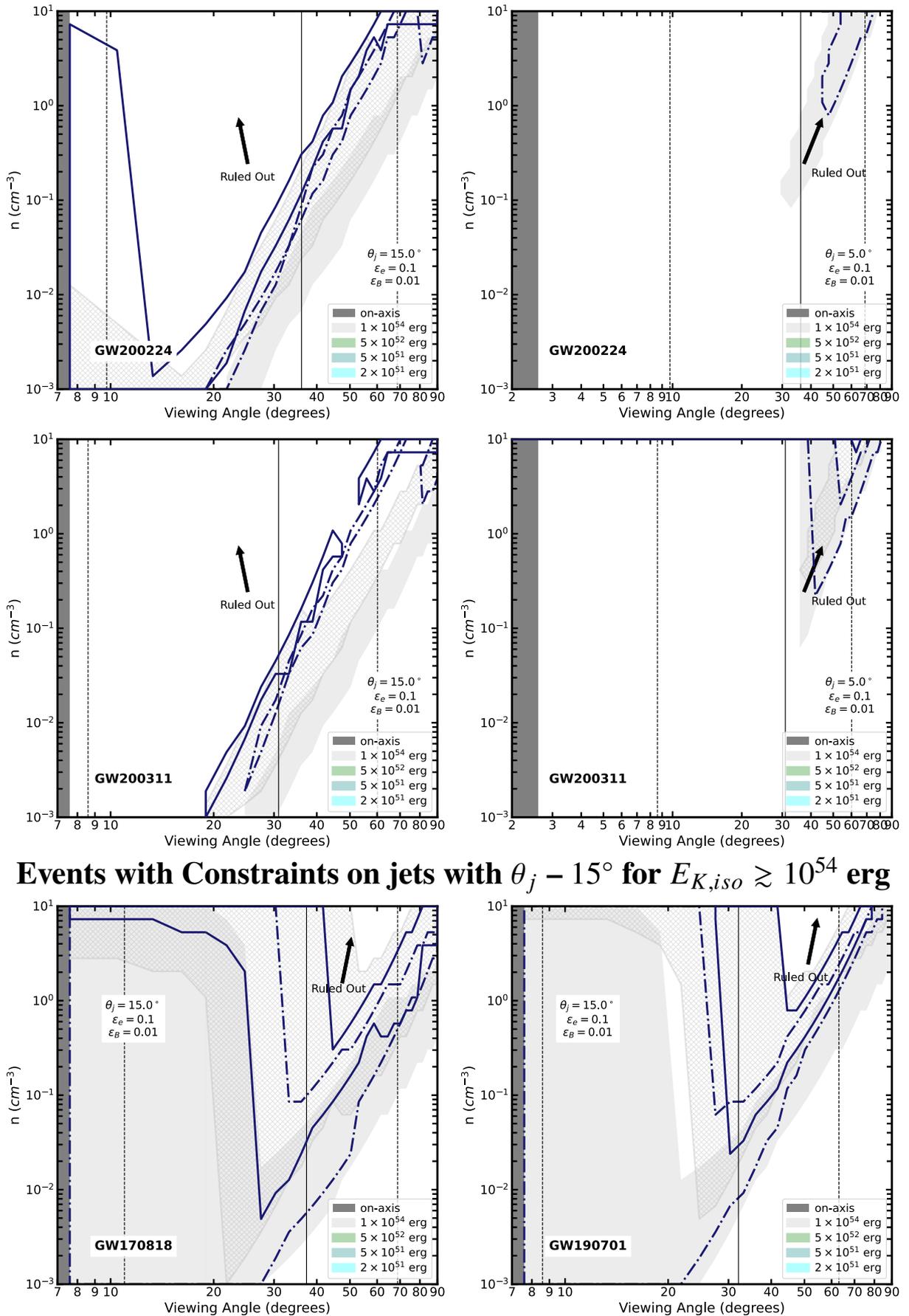

**Figure 6.** Observational constraints for events GW200224, GW200311, GW170818 and GW190701. Other details in Fig. 3 of the main paper.

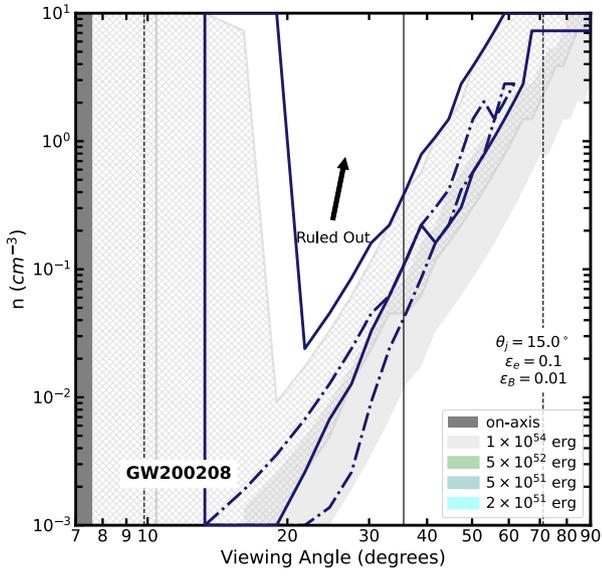

**Figure 7.** Observational constraints for event GW200208, if not treated as a candidate. Other details in Fig. 3 of the main paper.

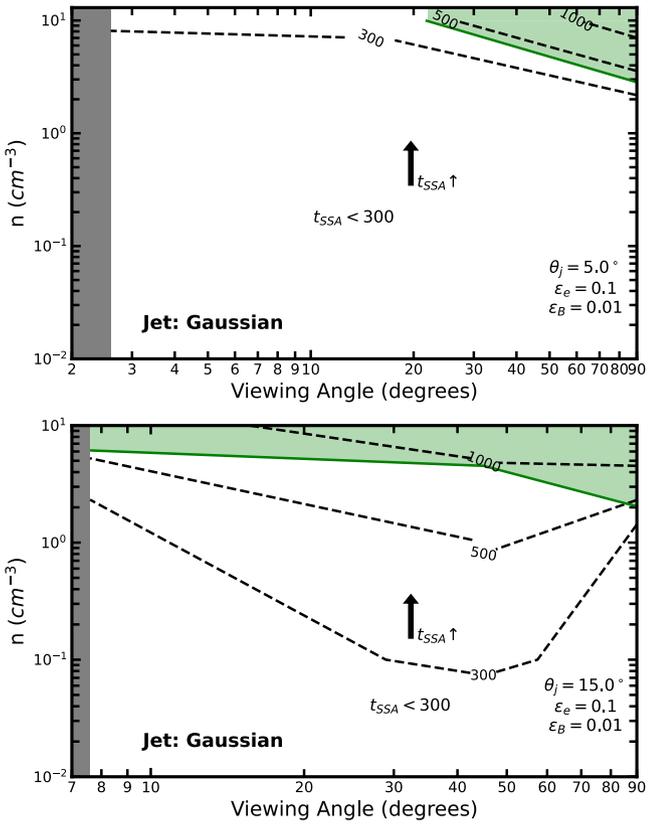

**Figure 8.** The number of days that an afterglow is self-absorbed for off-axis afterglows of Gaussian jets with $\theta_j$ of 5° and 15°, corresponding to top and bottom panels respectively. Other details in Fig. 6 of the main paper.